\documentclass[]{aa}
\bibpunct{(}{)}{;}{a}{}{,} 

\usepackage{graphicx}

\usepackage{txfonts}
\usepackage{hyperref}

\usepackage{soul}
\usepackage{color}
\usepackage{lineno}

\begin{document}
\title{SCALA: In-situ calibration for \\ Integral Field Spectrographs} 

\author
{
 {S.~Lombardo}\inst{\ref{1}}
 \and {D.~K\"usters}\inst{\ref{1}}
 \and {M.~Kowalski}\inst{\ref{1},\ref{8}}
 \and {G.~Aldering}\inst{\ref{2}}
 \and {P.~Antilogus}\inst{\ref{3}}
 \and {S.~Bailey}\inst{\ref{2}}
 \and {C.~Baltay}\inst{\ref{4}}
 \and {K.~Barbary}\inst{\ref{2}}
 \and {D.~Baugh}\inst{\ref{14}}
 \and {S.~Bongard}\inst{\ref{3}} 
 \and {K.~Boone}\inst{\ref{2},\ref{7}}
 \and {C.~Buton}\inst{\ref{6}}
 \and {J.~Chen}\inst{\ref{14}}
 \and {N.~Chotard}\inst{\ref{6}}
 \and {Y.~Copin}\inst{\ref{6}}
 \and {S.~Dixon}\inst{\ref{2}}
 \and {P.~Fagrelius}\inst{\ref{2},\ref{7}}
 \and {U.~Feindt}\inst{\ref{18}}
 \and {D.~Fouchez}\inst{\ref{13}}
 \and {E.~Gangler}\inst{\ref{9}}
 \and {B.~Hayden}\inst{\ref{2}}
 \and {W.~Hillebrandt}\inst{\ref{16}}
 \and {A.~Hoffmann}\inst{\ref{21}}
 \and {A.~G.~Kim}\inst{\ref{2}}
 \and {P.-F.~Leget}\inst{\ref{9}}
 \and {L.~McKay}\inst{\ref{22}}
 \and {J.~Nordin}\inst{\ref{1}}
 \and {R.~Pain}\inst{\ref{3}}
 \and {E.~P\'econtal}\inst{\ref{12}}
 \and {R.~Pereira}\inst{\ref{6}}
 \and {S.~Perlmutter}\inst{\ref{2},\ref{7}}
 \and {D.~Rabinowitz}\inst{\ref{4}}
 \and {K.~Reif}\inst{\ref{23}}
 \and {M.~Rigault}\inst{\ref{1}}
 \and {D.~Rubin}\inst{\ref{2},\ref{17}}
 \and {K.~Runge}\inst{\ref{2}}
 \and {C.~Saunders}\inst{\ref{3}}
 \and {G.~Smadja}\inst{\ref{6}}
 \and {N.~Suzuki}\inst{\ref{2}, \ref{19}}
 \and {S.~Taubenberger}\inst{\ref{16}, \ref{20}}
 \and {C.~Tao}\inst{\ref{13},\ref{14}}
 \and  {R.~C.~Thomas}\inst{\ref{10}}
 \\ (The Nearby Supernova Factory)
}

\institute{
    Institut f\"ur Physik, Humboldt-Universit\"at zu Berlin,
    Newtonstra\ss e 15, 12489 Berlin, Germany\label{1}
 \and
    Physics Division, Lawrence Berkeley National Laboratory, 
    1 Cyclotron Road, Berkeley, CA, 94720\label{2}
\and
    Laboratoire de Physique Nucl\'eaire et des Hautes \'Energies,
    Universit\'e Pierre et Marie Curie Paris 6, Universit\'e Paris Diderot Paris 7, CNRS-IN2P3, 
    4 place Jussieu, 75252 Paris Cedex 05, France\label{3}
 \and
    Department of Physics, Yale University, 
    New Haven, CT, 06250-8121\label{4}
 \and
    Universit\'e de Lyon, F-69622, Lyon, France ; Universit\'e de Lyon 1, Villeurbanne ; 
    CNRS/IN2P3, Institut de Physique Nucl\'eaire de Lyon.\label{6}
 \and
    Department of Physics, University of California Berkeley,
    366 LeConte Hall MC 7300, Berkeley, CA, 94720-7300\label{7}
 \and
    Deutsches Elektronen-Synchrotron, D-15735 Zeuthen, Germany\label{8}
 \and
    Clermont Universit\'e, Universit\'e Blaise Pascal, CNRS/IN2P3, Laboratoire de Physique Corpusculaire,
    BP 10448, F-63000 Clermont-Ferrand, France\label{9}
 \and
    Computational Cosmology Center, Computational Research Division, Lawrence Berkeley National Laboratory, 
    1 Cyclotron Road MS 50B-4206, Berkeley, CA, 94720\label{10}
 \and
    Centre de Recherche Astronomique de Lyon, Universit\'e Lyon 1,
    9 Avenue Charles Andr\'e, 69561 Saint Genis Laval Cedex, France\label{12}
 \and
   Aix Marseille Université, CNRS/IN2P3, CPPM UMR 7346, 13288, Marseille, France\label{13}
 \and
    Tsinghua Center for Astrophysics, Tsinghua University, Beijing 100084, China\label{14}
 \and
    Max-Planck-Institut f\"ur Astrophysik, Karl-Schwarzschild-Str. 1,
    85741 Garching bei M\"unchen, Germany\label{16}
 \and 
    Physikalisches Institut, Nussallee 12, Universit\"at Bonn, Bonn, Germany\label{21}
 \and
    Space Telescope Science Institute, 3700 San Martin Drive, Baltimore, MD 21218\label{17}
 \and 
 The Oskar Klein Centre, Department of Physics, AlbaNova, Stockholm University, SE-106 91 Stockholm, Sweden\label{18}
 \and 
    Institute for Astronomy 640 North A'oh$\mathrm{\bar{o}}$k$\mathrm{\bar{u}}$ Place, \#209 Hilo, HI96720-2700 USA\label{22}
 \and
    Bonn-Shutter UG, Auf dem H\"ugel 71, Universit\"at Bonn, Bonn, Germany\label{23}
 \and
 Kavli Institute for the Physics and Mathematics of the Universe, University of Tokyo, 5-1-5 Kashiwanoha, Kashiwa, Chiba, 277-8583, Japan\label{19}
   \and 
 European Southern Observatory, Karl-Schwarzschild-Str. 2, 85748 Garching, Germany\label{20}
}

\abstract
{}
{The scientific yield of current and future optical surveys is increasingly limited by systematic uncertainties in the flux calibration. This is the case for Type Ia supernova (SN~Ia) cosmology programs, where an improved calibration directly translates into improved cosmological constraints. Current methodology rests on models of stars. Here we aim to obtain flux calibration that is traceable to state-of-the-art detector-based calibration.}
{We present the SNIFS Calibration Apparatus (SCALA), a color (relative) flux calibration system developed for the
SuperNova Integral Field Spectrograph (SNIFS), operating at the
University of Hawaii 2.2$\,$m (UH\,88) telescope.}
{By comparing the color trend of the illumination generated by SCALA during two commissioning runs, and to previous laboratory measurements, we show that we can determine the light emitted by SCALA with a long-term repeatability better than 1\%.
We describe the calibration procedure necessary to control for system aging.
We present measurements of the SNIFS throughput as estimated by SCALA observations.}
{The SCALA calibration unit is now fully deployed at the UH\,88 telescope, and with it color-calibration between 4000\,{\AA} and 9000\,{\AA} is stable at the percent level over a one-year baseline.}

\keywords{telescopes - instrumentation: miscellaneous - standards - methods: data analysis}

\maketitle

\clearpage 
\section{Introduction}
\label{sec:intro} 

The ability to convert instrumental signals onto a physical flux scale has long been important for astrophysical applications. However, the precision and accuracy demanded has become increasingly stringent,
particularly for modern cosmology. The use of the distance-redshift relation of Type Ia supernovae (SNe~Ia) to derive the properties of the Universe, such as the dark energy equation of state parameter $w$, exemplifies this issue. SN~Ia cosmology compares the luminosity of a given restframe wavelength region (usually the Johnson B-band) of the SN~Ia spectral energy distribution (SED), at various redshifts. Thus, an accurate ``absolute color'' calibration is what matters most for cosmology.

The earliest efforts at optical wavelengths consisted of comparing the Sun, or bright stars such as Vega, to calibrated light sources placed at a distance. These light sources were either laboratory-calibrated lamps \citep{SK57, Hayes} or blackbody sources set-up in situ \citep{Oke, Tug77}. 
Such calibrations have quoted accuracies around 2\%. An alternative route has been to use theoretical models of stellar spectral energy distributions (SEDs). The most successful of these efforts has used three hot DA white dwarfs located in the Local Bubble and therefore essentially free of extinction by dust \citep{bohlin, bohlin_2014, rauch}. 
This model-based system has high internal accuracy as determined by the consistency between these white dwarfs, but its absolute color accuracy is difficult to evaluate independently of stellar models. Any error in the model of the color of these standard stars will directly propagate to the flux calibration applied to the SN fluxes. 
Flux calibrations from both of these systems have been transferred to observations of secondary standard stars. Observations of an extended network of primary and secondary standard stars on the model-based system, known as CALSPEC\footnote{\url{http://www.stsci.edu/hst/observatory/cdbs/calspec.html}} \citep{bohlin_2014}, provide the calibration of the instruments mounted on HST.

There are ongoing efforts to develop new techniques and instruments for physical calibration, especially for absolute color calibration. These new approaches focus on using precision detector-based laboratory calibration to monitor a light source used to illuminate a telescope and its attached science instrument. The National Institute of Standards and Technology (NIST) has found such dectector-based calibration to be more reliable than the older system based on calibrated light sources. The basic method of such systems consists of sequentially observing several quasi-monochromatic light sources with a telescope and then comparing the amount of light measured by the instrument with a calibrated detector that continuously monitors the emitted light. Examples of suitable light sources include tunable lasers, monochromators or LEDs. At optical wavelengths, NIST-calibrated silicon photodiodes are commonly used as the detector.

Once such a flux calibration is performed for a spectroscopic instrument, relevant standard stars can be directly observed. From these, after measuring and correcting for atmospheric extinction, a flux calibration system independent of the white dwarf models and the CALSPEC ladder could be constructed. For example, NIST has developed the Telescope Calibration Facility (TCF), whereby small telescopes are calibrated and then transported to an observatory to measure standard stars. The NISTstars program \citep{mcgraw_2012} plans to use such a TCF-calibrated telescope to transfer NIST calibration to standard stars. The ACCESS mission will calibrate its telescope before and after a sub-orbital flight that will observe several standard stars \citep{access_2016}. Our SNIFS Calibration Apparatus calibrates our telescope+instrument system in situ.

Beyond the establishment of an absolute color calibration, existing SN~Ia cosmology data consist of imaging through $\sim5$ filters whose responses as a function of wavelength must also be established in order to transfer absolute color calibration to SNe at different redshifts. The mapping of standard star fluxes to SNe when using filters is not unique; the signals to be compared are integrals of standard star or SN fluxes over a bandpass. SNe have complex spectra over the wavelength ranges spanned by broadband filters. In addition, bandpasses may change with time. Furthermore, any ground-based instrument that uses these standards will also need to correctly account for atmospheric extinction across their bandpasses. That low- and high-redshift SNe are usually observed with completely different telescopes, instruments, and filter sets adds to the difficulty. Thus, it is not surprising that measurements based on current SN~Ia cosmology samples are dominated by calibration systematic uncertainties, directly impacting the error on the measured value of $w$ \citep{betoule_2013,Betoule}.

A great deal has already been learned about how to approach these calibration challenges from efforts to calibrate imaging systems for SN~Ia cosmology. Such systems have generally focused on in-situ measurements of the passband shapes in order to obtain calibration using existing standard stars. A common configuration consists of a diffusive screen illuminated by a lamp-monochromator system or tunable laser \citep{stubbs2, stubbs3, stritzinger_2011, Marshall}. While illuminating the entire entrance pupil at once, with sufficient light to keep the calibration time short enough, the amount of stray light produced is often very difficult to correct for. This, along with screen non-uniformity, constitutes the limiting factor of these systems \citep{stubbs3}.

Other system configurations have been implemented or are planned. PanSTARRS switched from a diffusive screen to a sub-pupil projector system illuminating 0.2\% of the primary mirror \citep{tonry_2012}. \cite{Dice} direct a quasi-parallel sub-pupil beam of LED light onto the MegaCam imager. They find that the extended and diverging beam generate chromatic ghosts whose removal is challenging. LSST plans to use the combination of a diffusive screen and a projector beam to allow for large-scale and point-like corrections \citep{couglin_2016}. The ALTAIR project \citep{Albert} employs the more radical approach of flying calibrated laser diodes on a balloon (or even a satellite) that can be observed by a telescope. Such observations will include atmospheric extinction between the telescope and balloon, which must be determined separately by scanning the source over a large airmass range or measuring the extinction some other way.

However, even if these imaging systems were to deliver calibrated observations of CALSPEC stars, that calibration would only be valid for that instrument at that time. Beyond that, differences in bandpasses between instruments, field-angle dependence across a given filter, and the likelihood of throughput variations with time, limit the general applicability of such broadband imaging calibration systems. This is why we have focused on a spectroscopic approach.

Here we describe first results of a new calibration approach using the SNIFS Calibration Apparatus SCALA, first described in \cite{Lombardo}. 
SCALA is an in-situ flux-calibration device developed for the SuperNova Integral Field Spectrograph \citep[SNIFS,][]{lantz},  mounted on the University of Hawaii 2.2$\,$m telescope on Mauna Kea.  
 The ultimate purpose of SCALA is to calibrate the instrumental response of the ``$\mathrm{telescope}+ \mathrm{SNIFS}$''  system to 1\% precision. 
The advantage in calibrating SNIFS with such precision is that it is an integral field spectrograph, thus allowing us to produce spectrophotometric calibrated spectra of standard stars.
These spectra are also corrected for the  atmospheric extinction which is computed nightly from standard stars observations \citep{buton_atmospheric_2013}.
Therefore, the resulting physical calibration can be directly transferred to standard stars. 
These can then be used in a completely general fashion by any telescope, whether in space or on the ground.
Additional advantages of SCALA are the high elevation of Mauna Kea, resulting in less overall extinction, low absorption due to atmospheric water, and the existence of a well-defined inversion layer that further suppressed extinction from aerosols -- the most time-variable broadband atmospheric extinction component.

In the next sections we discuss the concept and design of SCALA (Section~\ref{sec:concept}), and the data obtained using its calibrated photodiodes (Section~\ref{sec:clap_datataking}).
In Section~\ref{sec:control} we discuss in detail the key pre-commissioning and post-commissioning tests conducted to measure the performance and stability of the system. The second post-commissioning run additionally commissioned an aperture mask; comparisons with and without the mask, and measurements of standard stars are discussed.
We explain the calibration methodology that was implemented, and show the first throughput measurements in Section~\ref{sec:strategy} and \ref{sec:data_take}. The origins of, and limits on, potential sources of systematic uncertainty are discussed in Section~\ref{sec:systematics}. Some of the SCALA systematic uncertainties have already been described in detail in \cite{kuesters}, hereafter K16.
Finally, we discuss some ancillary uses for SCALA in Section~\ref{sec:potential_use} and then conclude in Section~\ref{sec:conclusion}.

\section{The SCALA concept and design}
\label{sec:concept} 

Our goal for SCALA is twofold: 
to first calibrate the ``$\mathrm{telescope}+\mathrm{instrument}$'' system response at the percent level, and second, to use this calibration to redefine the standard star network. 
SCALA has been built in close collaboration with the Nearby Supernova Factory \citep{aldering}. 
Here, we will focus on relative flux calibration, which is the part of fundamental calibration that impacts SN~Ia cosmological analyses.

SCALA is an in-situ flux-calibration device designed to produce a uniform illumination of the focal plane, i.e., a flat-field, while monitoring the light emitted using a calibrated photodiode.
An ideal calibration source would generate a parallel light beam illuminating the entire entrance pupil of the telescope in order to mimic the optical path of the calibration target, usually a star. 
This would require a point-source-like object at infinity as the light source, which is difficult to build and control at the desired precision, or a full-aperture collimator, which would not be practical for such a large telescope. 
While a diffusive screen also illuminates the entire entrance pupil, the resulting light path is intrinsically different from that of a star.
Our strategy with SCALA was to build an ``artificial planet'' of angular size $1^\circ$. 
Even though it is extended, it shares most of the characteristics of a point source, e.g. a collimated beam.
In the following subsections we briefly describe SNIFS and the subsequent constraints on the SCALA design.
 
\subsection{SNIFS}
\label{sec:snifs}
The SuperNova Integral Field Spectrograph (IFS) SNIFS \citep{lantz}, located at the bent Cassegrain port of the UH 2.2\,m on the summit of Mauna Kea, is composed of three channels: two spectroscopic (blue and red, respectively) and one imaging. Each spectrograph camera holds a E2V 2k$\times$4k CCD. 
The dual-channel spectrograph simultaneously covers 3200--5200\,{\AA} ($B$-channel) and 5100--10\,000\,{\AA} ($R$-channel) with  line spread functions having FWHM values of 5.23\,{\AA} and 7.23\,{\AA}, respectively. The spectrograph samples a $6.4"\times6.4"$ field-of-view through a combination of a microlens array made of 15$\times$15 lenses followed by a collimator, a grism and a camera. Each of the 225 $0.43"\times0.43"$ spatial elements of the microlens arrays that segment the focal plane is called a spaxel.

The imaging channel covers a $9.4'\times9.4'$ field-of-view, covered by two 2k\,$\times$\,4k E2V CCDs. 
It is equipped with a filter wheel composed of $ugriz$ (SDSS) filters, a pinhole grid and a multiple-band filter.
In regular SNIFS observations, the filter wheel is set on the multiple-band filter and the images are used for guiding and relative flux calibration during non-photometric nights \citep{pereira_tesi}. 
The multiple-band filter produces observations of different areas of the sky in distinct bands.
 As discussed in Sections~\ref{sec:alignment} and \ref{sec:systematics}, we use the imaging channel for aligning the telescope with SCALA and for testing potential sources of systematic uncertainty in the SCALA system.

SNIFS data reduction was summarized by \cite{aldering_2006} and updated in Section~2.1 of \cite{scalzo}.  
The flux calibration was developed in Section~2.2 of \cite{pereira_2013}, based on the atmospheric extinction derived in \cite{buton_atmospheric_2013}.
The SNIFS data used in this paper are reduced according to the initial steps of SNIFS pipeline, e.g. bias removal and wavelength calibration, but then stopping the pipeline reduction before flat fielding and flux calibration.
By processing our data in this way we can reduce the possible systematic difference between the processing of calibration and science frames.

\subsection{Constraints for SCALA}
\label{sec:scala_constrains}

The goal of using SCALA is to derive an accurate UH\,88+SNIFS flux calibration and, using the \cite{buton_atmospheric_2013} atmospheric modeling procedure, to recalibrate the standard
star network. This sets constraints on the 
design of SCALA and its associated observation strategies. 
The requirements for SCALA are such that it must:

\begin{enumerate}
 
  \item be tunable over the 3200--10000\,{\AA} range, to ensure coverage of the full SNIFS spectral range.
  \item continuously monitor the output light in order to check for intensity variations or irregularities. 
  \item illuminate the entire entrance pupil or at least a representative sampling of it, to catch possible large-scale achromatic illumination gradients.
  \item present uniform illumination of a field of view larger than $9'\times9'$ in order to calibrate the imaging channel and its filters as well.
  \item be mounted to the telescope dome, so it is accessible on demand.
  \item be mounted at an elevation that is within the range of normal science operations in order to have the ability to account for the gravitational load on the telescope optics + baffling that could, in principle, affect flux calibration.
  \item be fully controllable remotely, so it can be used without personnel on site.
\end{enumerate}

These characteristics brought us to the design described in the following sections.
\subsection{SCALA Design}
\label{sec:design} 

\begin{figure}
\centering
\includegraphics[scale=0.21]{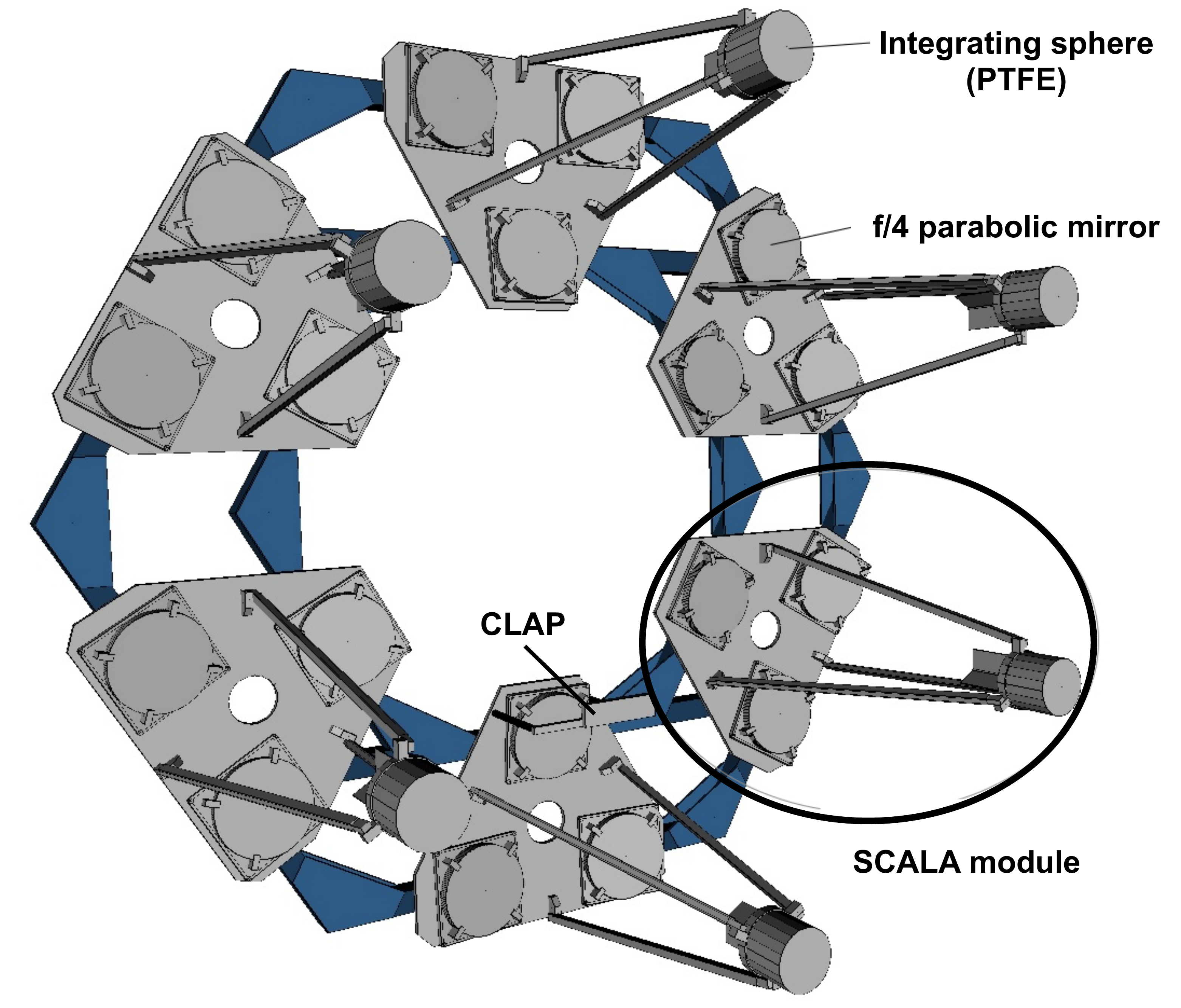}
\includegraphics[scale=0.25]{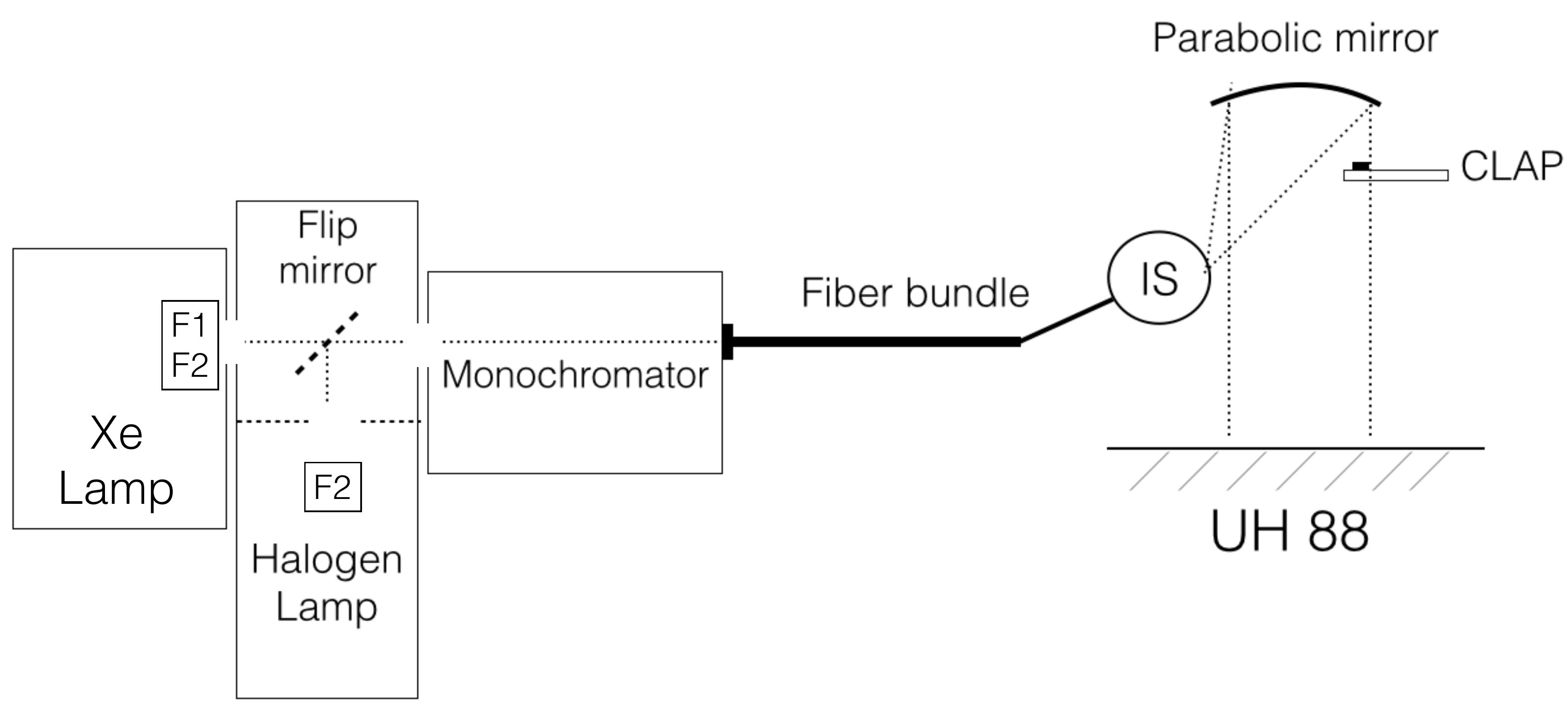}
\caption{Top: hexagonal arrangement of the six submodules of SCALA. This structure is mounted in front of the entrance pupil of the telescope. The lower module shows the position of the reference photodiode (CLAP). Bottom: SCALA scheme where it is possible to see the lamp system and the flip mirror that allows the selection of lamp output that illuminates the monochromator entrance, the fiber bundle that feeds the integrating sphere (IS), and the calibrated photodiode (CLAP), which faces the collimated beam reflected by the mirror. The dotted line represents the light path. The flipping mirror is placed in the housing of the halogen lamp to avoid light leaks. The two small boxes within the lamps boxes represent the place where the order sorting filters (F1 and F2) are located. In this schematic drawing, the angle between the optical axes of the IS and mirror is exaggerated and the IS output is simplified to a point source.}
\label{Flo:hexagon}
\end{figure}

SCALA consists of 18 mirrors, whose beams are distributed over the UH\,88 entrance pupil in a nearly hexagonal arrangement pointing at the telescope (see Figure~\ref{Flo:hexagon}, top panel). 
Integrating spheres fed by a lamp-monochromator system illuminate the mirrors, producing 18 collimated beams. Each beam has an opening angle of $1^\circ$, thereby forming a uniform "planet" rather than a star.

The stability and the flux intensity produced by the light source are monitored by a Cooled Large Area Photodiode \citep[CLAP,][]{Dice} that faces one of the f/4 mirrors and, there, continuously measures the light output.
Because the light output from SCALA does not illuminate the entire UH\,88 mirror, a pupil mask is mounted at the top of the telescope during SCALA calibration campaigns. 
With the mask mounted, the region of the primary mirror covered by standard star light will be exactly the same as that calibrated by SCALA.
Note that the mask is only necessary when recalibrating standard stars -- wavelength-relative throughput measurements can be made without it.
A conceptual diagram summarizing the layout of a single mirror module is shown in the bottom half of Figure~\ref{Flo:hexagon}.

A more detailed description of each of the SCALA components --~the lamp system, the projector modules, the light monitoring system, and the mask~-- is provided in the following subsections. The system control software is presented in Section~\ref{sec:scala_soft}, and the technique used to align the SCALA optics to the telescope optical axis is described in Section~\ref{sec:alignment}.

\subsubsection{The lamp system}
\label{sec:design_lightsource}

We illuminate a Newport Cornerstone 260 monochromator equipped with two gratings, ruled at 1200~l/mm and having blaze wavelengths of 3500\,{\AA} and 7500\,{\AA} respectively. This combination delivers tunable light ranging from 3200\,{\AA} to 10000\,{\AA}. 
The line spread function of SCALA is 35\,{\AA} FWHM, and is set by the dimensions of the entrance and exit slits of the monochromator. It was selected to have a good balance between spectral resolution and light level.
The monochromator is fed by two different lamps, depending on the desired wavelengths, in order to provide an emission-line free source. 
An APEX-Illuminator with 150\,W Xe lamp is used for the 3200-7020\,{\AA} wavelengths and a halogen lamp is used for redder wavelengths.

The lamp system is also equipped with order sorting filters. Filter-type F1 cuts on at 3090\,{\AA} and filter-type F2 cuts on at 4950\,{\AA}. These are used to prevent second order light at the output of the monochromator when using the 3500\,{\AA} or 7500\,{\AA} blaze gratings, respectively. 
Both F1- and F2-type filters are located on a filter wheel at the exit port of the Xe lamp, while a F2-type filter is mounted on the optical path followed by the light in the halogen lamp, as illustrated in the bottom half of Figure~\ref{Flo:hexagon}.
Weak second order light starts to appear at 9000\,{\AA} when using a F2-type filter, but remains subdominant until 9700\,{\AA} ($<0.8$\% with respect to the light emitted at the requested wavelength). 
However, here we conservatively limit our calibration to wavelengths bluer than 9000\,{\AA}.
Usage of an order sorting filter with a redder cut-on than F2 will be necessary in order to suppress second-order light out to the wavelength limit of the CCDs.
In Table~\ref{tab: list_setup} the different configurations of lamps, gratings and filters used in the wavelength range during regular calibration, are listed. For gratings and filters, the values reported are the blaze and cut-on wavelengths respectively.

\begin{table}[h]
\begin{center}
\caption{Configurations of lamps, gratings and filters, used when operating SCALA for different wavelength ranges.}
\begin{tabular}{cccc}
\hline\\[-1.8ex]
Wavelength    & Lamp    & Grating   & Filter \\
Range         &         & blaze     & cut-on  \\
 \,[{\AA}]    &         & \,[{\AA}] & \,[{\AA}] \\
\hline\\[-1.8ex]
3200 -- 4500  & Xe      & 3500      & none \\[0.25ex] 
4500 -- 5220  & Xe      & 3500      & 3090 (F1)\\ [0.25ex]
5220 -- 6240  & Xe      & 7500      & 3090 (F1)\\ [0.25ex]
6240 -- 7020  & Xe      & 7500      & 4950 (F2)\\ [0.25ex]
7020 -- 10000 & halogen & 7500      & 4950 (F2)\\ [0.25ex]
\hline\\[-1.1ex]
\end{tabular}
\label{tab: list_setup}
\end{center}
\end{table}

\subsubsection{The projector modules}
\label{sec:design_modules}
The monochromatic light produced by the lamps and monochromator system is transferred via optical fibers to six individual modules which, in turn, project the light into the UH\,88. 
Each of the modules is composed of an integrating sphere (IS) and three 20~cm diameter f/4 parabolic mirrors. 
Each IS is fed by a fiber bundle (Ceramoptec, Optran WF) and has three 1.4~cm diameter opening holes facing each of the mirrors.
This novel integrating sphere concept designed for SCALA enables a uniform and well 
collimated beam, as detailed in \cite{Lombardo}, based on the concept developed by \cite{vaz_thesis}\footnote{\url{http://hdl.handle.net/1721.1/65435}}. 
The fiber bundle guiding the light from the monochromator to the individual modules is structured such 
that each of its six arms has fibers distributed across the exit slit of the monochromator 
to guarantee a homogeneous sampling of the entire slit across integrating spheres.

As illustrated in Figure~\ref{Flo:hexagon}, the six modules are mounted in a hexagonal configuration.
The beam intensity gradient, caused by the off-axis mirror systems, is canceled by the symmetry of the system -- both of the modules and the hexagonal configuration -- as described in \cite{Lombardo}.
  
\subsubsection{The monitoring system}
\label{sec:design_clap}

The reference system continuously monitors the light that is directed into the telescope.
For this task we use two Cooled Large Area Photodiodes (CLAPs).
These have been directly calibrated to a NIST-calibrated photodiode.
The two CLAPs have been produced by the DICE team \citep{Dice} and used also by their calibration experiments, SnDICE and SkyDICE.
These photodiodes (Hamamatsu S3477-04) are very stable and highly sensitive ($0.5$\,A/W at 9600\,{\AA}).
Their calibration precision is 0.7\% or better in the wavelength range of interest.
They have a sensitive area of $5.8\times5.8$\,mm$^2$ and are equipped with a two-stage Peltier cooler, which keeps them at a temperature of $-14\,^{\circ}$C.

The front-end board of each CLAP includes the photodiode and a low-noise amplifier, while the signal digitization is performed on the back-end. For our measurements we set the CLAP sampling frequency at 1\,kHz. These devices are small and compact enough to be held by a U-shaped structure (as illustrated in Figure~\ref{Flo:hexagon}) that is mounted in front of our beams without obscuring the light from the integrating sphere. 
This structure allows their installation in front of any SCALA projection mirror.

During operations, one photodiode is used to monitor the light emitted by SCALA. The second photodiode is used for tests and serves as second monitor of the system (further described in the following sections).

\subsubsection{The mask}
\label{sec:mask}
The 18-mirror design of SCALA illuminates $17\%$ of the primary mirror. 
We designed a mask to cover the non-illuminated regions.
This mask is used both when calibrating the ``$\mathrm{telescope}+ \mathrm{SNIFS}$'' system's response 
and when observing the standard stars for the purposes of calibrating them (this will be further discussed in future analyses).
In this way, the effective optical path of light from celestial targets will be calibrated by SCALA.

The mask is mounted at the top of the UH\,88 telescope. It is made of ALUCORE painted with a matt black finish and has holes 16~cm in diameter, aligned to match the SCALA beams (as detailed in K16).
These holes are undersized to allow an alignment tolerance of $\pm2$\,cm transverse to the beams.
As a result, when the mask is in place, SCALA or standard stars illuminate 10\% of the primary mirror.
 
\subsubsection{The software}
\label{sec:scala_soft}
Except for the pupil mask, which needs to be mounted and unmounted manually, 
the entire SCALA system is automatic and remotely controllable. 
Monochromator, lamps, photodiodes, and a NetIO 230C, used to switch the other SCALA 
elements on and off, are connected to a computer.

The software controlling SCALA and the CLAP data acquisition system has been written in Python 
-- the code is available online \footnote{\url{https://github.com/snfactory/scala}}. 
The user can set a list of wavelengths and exposure times and the software will
automatically set the SCALA configuration, i.e., the monochromator, grating, filter and lamp.
In addition, we have implemented a SNIFS--SCALA interface such that the SNIFS control software can directly 
control SCALA and observe it as if it were any other astronomical target (see further details in Section~\ref{sec:strategy}).
The flexible and modular design of SCALA means it could easily be adapted to other telescope+instrument combinations.

\subsubsection{Alignment procedure}
\label{sec:alignment}
In order to establish the correct alignment of all the SCALA optics with the telescope, we mounted an LED light source at the direct Cassegrain focal plane of the telescope and used it and an optical module to illuminate SCALA with a collimated, parallel beam. 
Each SCALA mirror was then fine-tuned by focusing this beam onto the center of the shutter in front of each IS hole. 
These shutters, which can be manually opened and closed to block the light from one hole, have spots painted on the location of the center to facilitate the alignment.

The alignment of the SCALA optics has been performed in both the 2014 and 2015 commissioning phases, with no measurable shift or flexure during the intervening time, except for an overall tilt of the entire structure due to a slightly loose screw, which set the tilt of the SCALA mounting with respect to the dome. The structure is therefore very stable.

The precise pointing of the telescope towards the SCALA direction is automatically performed using a routine whereby the telescope points in the general direction of SCALA and acquires a white-light image using the pinhole grid of the SNIFS imaging channel. 
In this way several images of the entrance pupil of the telescope with a SCALA-like pattern are created on the imaging CCDs.
Possible non-reproducibilities of the dome position will lead to position shifts 
of these patterns. 
Comparison with a reference image, for which SCALA was correctly aligned with the telescope, allows for a refinement of the telescope pointing under software control and the compensation of any dome position offset.

\section{Photodiode data}
\label{sec:clap_datataking}

The data obtained from the photodiode, used to monitor the SCALA light level, consist of three parts: (1) a two second background exposure, (2) a "light-on" exposure of variable length and (3) another two second background exposure.
The two background exposures, performed with the monochromator shutter closed, track the dark current of the photodiode and the background ambient light, and their potential time variation.

\begin{figure}[h]
\centering
\includegraphics[scale=0.23]{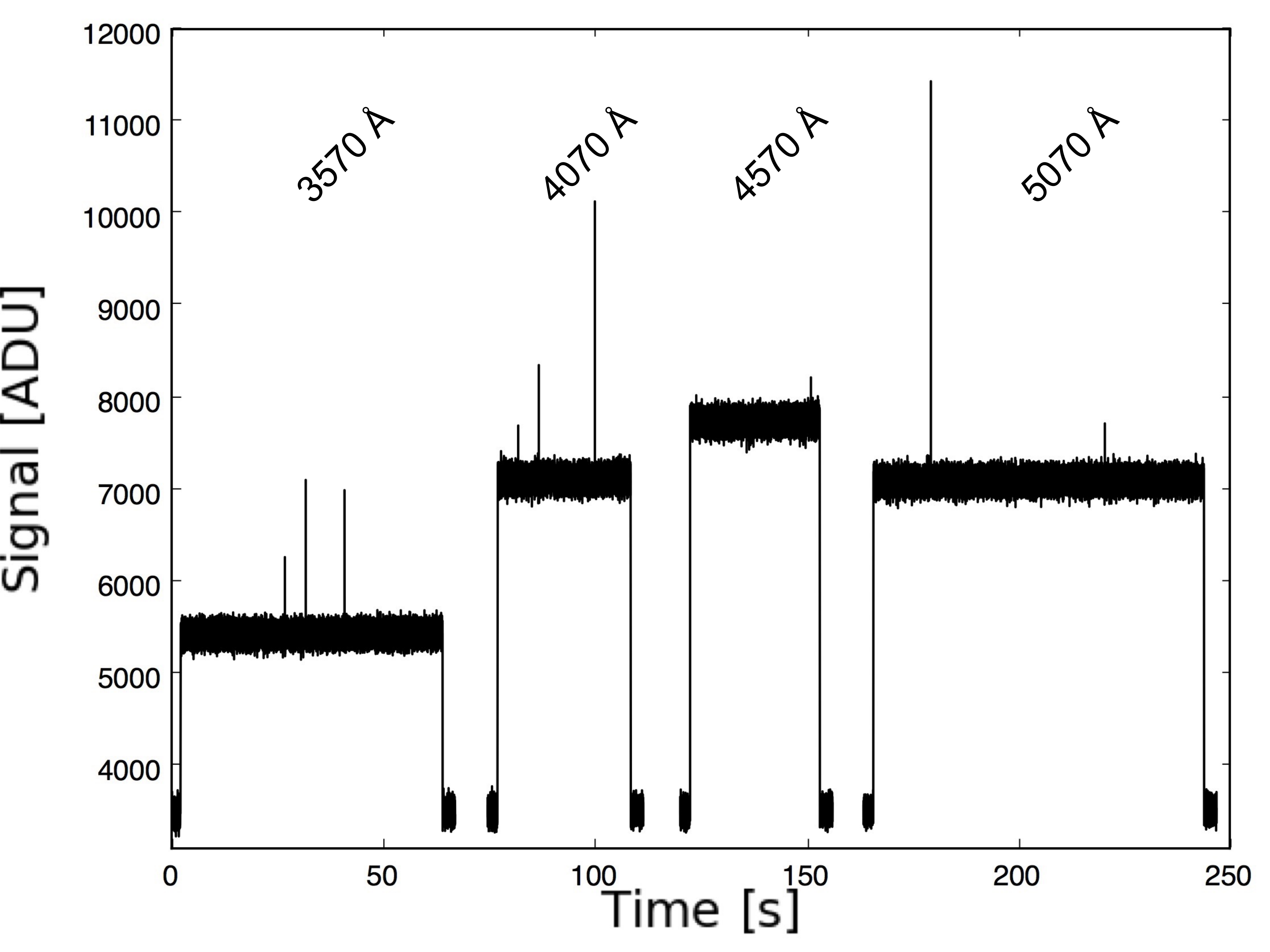}

\caption{Example of the data obtained from one CLAP for a SNIFS blue-channel exposure. Notice the different exposure time for each wavelength, depending on the light level and SNIFS transmission, and also the presence of $\sim2$\,s background exposures before and after observing each wavelength. The outliers in the data are cosmic rays, which are subsequently removed by the analysis software. The wavelengths observed are printed above the CLAP data for each exposure. }

\label{Flo:clap_data}
\end{figure}

As illustrated in Figure~\ref{Flo:clap_data}, the calibrated photodiode continuously 
records data for each wavelength exposure, with a frequency set at 1kHz. 
Laboratory tests were used to verify that the timescale for significant dark current variations is longer than the longest SCALA line exposure time, and that the dark current is smoothly varying.
An average of the two background measurements is therefore a good estimate of the dark current that the photodiode experienced during the light exposure.
This background averaging method achieves a random error $<0.2\%$ for calibration of most wavelengths for nighttime photodiode data.
The background subtraction procedure and its accuracy 
are further discussed in Section~\ref{sec:systematics} and in K16, where daytime ambient light  contamination is also discussed.

The shutter of the monochromator is the element ultimately determining the exposure time for each SCALA wavelength observation. 
CLAP data are not only used as a light monitor but also as independent means to measure the exposure time, with a precision better then $<0.01\%$ for our shortest exposure times.
With the 1~kHz sampling frequency of the CLAP, we are even able to measure the rising and falling ramp due to the opening and closure of the monochromator shutter.

\section{System characterization}
\label{sec:control}
The calibration of the ``$\mathrm{telescope}+ \mathrm{SNIFS}$'' system requires a precise characterization of our instrument. 
We structured this in two steps: first the measurement of the response of each SCALA component individually in the lab, and second, the measurement of the fully integrated system.
Consistency between the two approaches will verify that the SCALA output and monitoring is understood at the component level. Any detected differences would focus attention on the types of changes that may need careful monitoring or remeasurement when using SCALA in situ.
The two sets of measurements are detailed in Section~\ref{sec:control_individual} and 
Section~\ref{sec:control_scala}, respectively. 
In Section~\ref{sec:control_reproducibility} we discuss our ability to characterize the light output by SCALA, and how the light transmitted to the telescope by the 18 independent beams is scaled relative to the output of the beam monitored by the reference photodiode as function of wavelength.

\subsection{Response of SCALA components}
\label{sec:control_individual}

Laboratory measurements like those described in this section proved essential for selecting components able to produce consistency in the color output between the 18 SCALA beams. Good color uniformity is important because color differences between beams have the potential to introduce systematic errors if the reflectively of the primary mirror were also sufficiently non-uniform so as to reweight beams of different colors as seen by SNIFS.

Wavelength-relative throughput curves have been measured in a controlled laboratory environment for every single part of the system, including the mirrors, the ISs and the fiber bundle arms. These measurements used a monochromator and Xe lamp system separate from, but essentially identical to, the SCALA monochromator and Xe lamp.
\begin{figure*}
\centering
\includegraphics[scale=0.43]{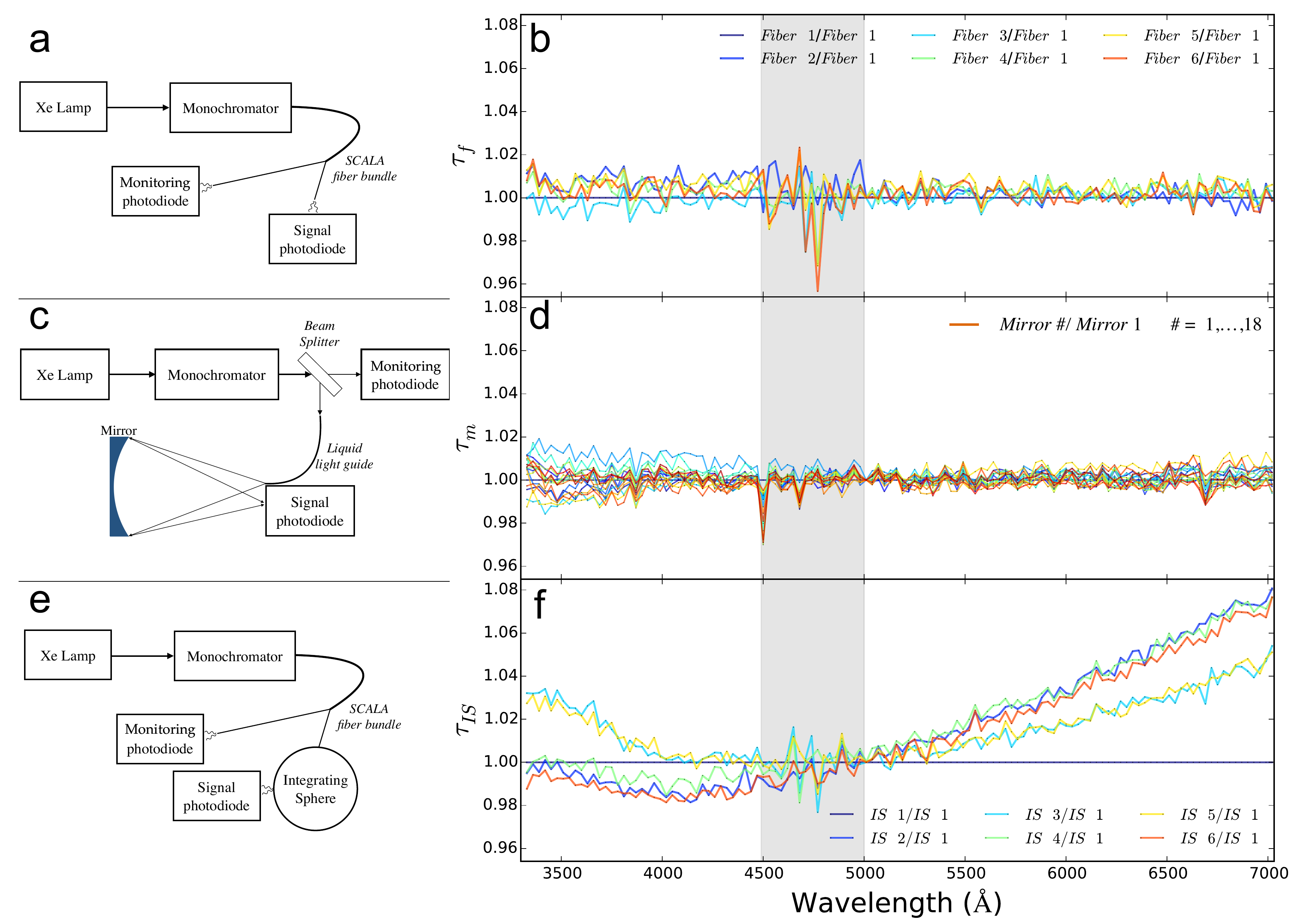}

\caption{Left (panels a, c, e): set-up used in the laboratory to measure the corresponding quantities on the right side of the plot. Right (panels b, d, f): relative responses of the SCALA fibers (b), mirrors (d), ISs (f), with respect to their reference component and normalized at 5000\,{\AA} for illustration purposes. The gray band delimits the region where the weak emission lines of the Xe lamp are located (4500-5000\,{\AA}).}
\label{Flo:bonn_test}
\end{figure*}
Two lock-in amplifiers were used in combination with a chopper to amplify the signal received by model UV-035 EQC (not CLAP) monitoring and signal photodiodes.
These measurements consist of ratios made using the same photodiode, at every wavelength, for both signal and monitoring systems and therefore the QEs of the photodiodes cancel out. 
We kept the monitoring setup fixed when measuring each type of SCALA component.
Since the small variations of the Xe lamp light level between measurements of the individual SCALA components are corrected by ratios of measurements made using the monitoring photodiode, the wavelength dependence of the component scanned by the monitoring setup also cancels out.
The reference components (labeled as IS 1, Fiber 1 and Mirror 1, in Figure~\ref{Flo:bonn_test}) were chosen to be those that were ultimately integrated to make the beam monitored by the reference CLAP when using SCALA in situ.

Our goal in the following sections is to compare the lab measurements with the commissioning measurements. However, the Xe lamp has bright emission lines longward of 7000\,{\AA}, and these are narrower than the resolution of the exit slit of the monochromator. These lines were found to cause too much variability in the comparison between different sets of laboratory measurements since even tiny variations in wavelength reproducibility of the monochromator grating led to strong changes in the monochromator output.
Therefore comparisons are possible only below 7000\,{\AA}.
The Xe lamp also has emission lines at 4500--5000\,{\AA}. However, as these lines are much weaker and introduce only modest added scatter (which we highlight in the relevant figures), the comparison data in this wavelength range can still be used.

\subsubsection{Optical fibers}
For the characterization of the optical fiber bundle arms, we implemented the measurements according to the setup shown in  Figure~\ref{Flo:bonn_test}a (upper left).
The input end of the SCALA fiber bundle was mounted at the exit of the monochromator. Then the end designated as the reference was monitored by the monitoring photodiode while the signal photodiode was used to measure the light from each of the remaining five fiber bundle arms. We cycled through the different fiber bundle arms until all five fiber bundle arms had been scanned in wavelength using a step of 30\,{\AA}.
These relative responses are plotted in Figure~\ref{Flo:bonn_test}b as ${\tau_{f}}$. They have been normalized at 5000\,{\AA} to aid in evaluating color trends.
As can be seen, the fiber bundle arms are very similar, having relative color trends smaller than 1\%.

\subsubsection{Mirrors}
The reflectivities of the 18 mirrors were measured using the same lamp and monochromator system as above, but this time a beam splitter divided the light between the monitoring photodiode (always the same) and the entrance of a liquid light guide.
The exit of this guide illuminated a SCALA mirror, and the light was then focused on the other photodiode, as shown schematically in Figure~\ref{Flo:bonn_test}c.
Each mirror was sequentially placed as shown in Figure~\ref{Flo:bonn_test}c and scanned in wavelength, without modifying the rest of the set-up. 
We scaled the signal photodiode measurements to the respective monitoring
photodiode signal to correct for lamp variation between the individual measurements.
In this way the wavelength trend of the unmodified part of the light path cancels out in the ratios.
In Figure~\ref{Flo:bonn_test}d we plot the ratios between the responses of the different mirrors with respect to that of the reference mirror as ${\tau_{m}}$, again, normalized at 5000\,{\AA}. 
The color trend in these reflectivity comparisons is mostly gray, except for the shortest wavelengths where it is around 1\%.

\subsubsection{Integrating spheres}
\label{sec:control_is}
Finally, the integrating sphere responses were measured. We first mounted the input end of the SCALA fiber bundle at the exit of the monochromator. 
Then the end of one fiber bundle arm was connected to the IS entrance port and the signal photodiode was positioned immediately in front of one of the IS exit ports (Figure~\ref{Flo:bonn_test}e).
The monitoring photodiode was used to simultaneously monitor the designated reference fiber bundle arm. For this series of measurements each IS was installed and measured in turn, without changing the rest of the set up.
Care was taken to fix the IS positions with respect to the signal photodiode, and to not change any other elements, while cycling through each IS.
In Figure~\ref{Flo:bonn_test}f the responses of the integrating spheres, ${\tau_{IS}}$, with respect to one of them are shown, again normalized at 5000\,{\AA}.
Here color trends of up to 8\% are observed. This is likely due to the slight composition differences of the PTFE (Teflon) blocks from which they were machined.
Fortunately, as we show in Section~\ref{sec:wave_uncertain} below, when we take this into account we see reproducible results, and minimal systematic uncertainty from these chromatic differences.

\subsection{Response of the fully integrated system}
\label{sec:control_scala}
The second step of the SCALA system characterization is a measurement of the relative throughput, as a function of wavelength, for each of the 18 beams in the assembled configuration. The precise characterization of the system can be ensured only if we know the behavior with wavelength of every SCALA beam. 

\subsubsection{Laboratory measurements}
In the previous subsections we showed that all the relative responses of SCALA components have different behaviors as functions of wavelength. 
The single component responses are now numerically combined to mimic the full system that was shown in Figure~\ref{Flo:hexagon} by calculating the quantity: 
\begin{equation}
 {P_{i,j,k}(\lambda)} = {{\tau_{f_i}(\lambda)}\cdot{\tau_{IS_j}(\lambda)}\cdot{\tau_{m_{k}}(\lambda)}}
\label{eq:module}
\end{equation} 
where $P_{i,j,k}(\lambda)$ is the relative response of one of the SCALA beams when using the $i$-th fiber bundle, $j$-th integrating sphere and $k$-th mirror, all taken with respect to the reference beam.
The $\tau$ functions in Equation~\ref{eq:module} are those plotted on the right side of Figure~\ref{Flo:bonn_test}, but in this case  without normalization at 5000\,{\AA}.
The sum over all 18 beams is used as an estimate of the total light expected to be produced by SCALA, and will be used in Section~\ref{sec:control_reproducibility}.

\subsubsection{In-situ measurements}
The light from SCALA as seen in the telescope focal plane consists of the sum of the flux reflected by the 18 mirrors.
Now we show how to construct the analog of Equation~\ref{eq:module} for SCALA installed at the observatory.
We express the wavelength dependence of the SCALA light reproducibility,
$E_{s}$, as follows:
\begin{equation}
 E_{s}({\lambda_{\mathrm{c}}}) = 1+\sum_{i=n=2}^{18}{\frac{C^{i(n)}_{\mathrm{t}} (\lambda_{\mathrm{c}})\cdot{D_\mathrm{t}(\lambda_{\mathrm{c}})}}{C^{n}_{\mathrm{r}} (\lambda_{\mathrm{c}})\cdot{D_\mathrm{r}(\lambda_{\mathrm{c}})}}},
\label{eq:a18}
\end{equation} 
where $C^{n}_\mathrm{r}$ is the $n$-th flux measurement by the reference CLAP corresponding to the flux, $C_\mathrm{t}^{i}$, measured 
by the test CLAP in the $i$-th beam for the SCALA line centered at $\lambda_{\mathrm{c}}$. $D_\mathrm{r}$ and $D_\mathrm{t}$ are the factors to convert the reference CLAP and the other CLAP measurements, respectively, from ADU to W, obtained by weighting the photodiode calibration curve -- provided by the DICE team -- by the SCALA line profile.
$E_{s}$ is therefore the number of times the light measured by the reference CLAP should be multiplied to represent the total SCALA output. 
If all beams were identical to the reference beam, $E_{s}$ would be equal to 18 for every wavelength.
Here, we neglect the fraction of beam obscured by the photodiode as it is always obscured by the primary mirror mask.

In order to measure the transmissions of all the beams for the 
in situ set-up we performed 17 independent sequences of SCALA monotonic scans in wavelength using one CLAP as reference 
-- always facing the same mirror -- and manually moving the other CLAP from one mirror to another.

In order to reflect a collimated beam without being obstructed by the IS, SCALA mirrors are mounted in an off-axis configuration.
Due to this geometry, there is an illumination gradient across the reflected beam. 
This means that the sensitive area of the photodiode experiences a different amount of light depending on which part of the beam it samples. 
Thus, we took care to position a CLAP to always cover the same region of the SCALA beam that was being measured. We were able to achieve a reproducibility better than 0.7\% in the wavelength range of interest (as will be shown in Section~\ref{sec:scala_area_stb}).
$E_{s}$  has been measured during the two commissioning phases in 2014 and 2015.

\subsection{SCALA light reproducibility}
\label{sec:control_reproducibility}
The comparison between the three sets of measurements of the total light produced by SCALA provides two important pieces of information regarding our system: how well it is understood at the component level and its stability over time. This rests on comparing the calculated sum over the components of the 18 beams from Equation~\ref{eq:module} to the quantity in Equation~\ref{eq:a18} measured in situ on the fully integrated system in 2014 and 2015.

\subsubsection{Lab measurements vs 2014 commissioning}
\begin{figure}[h]
\centering
\includegraphics[scale=0.25]{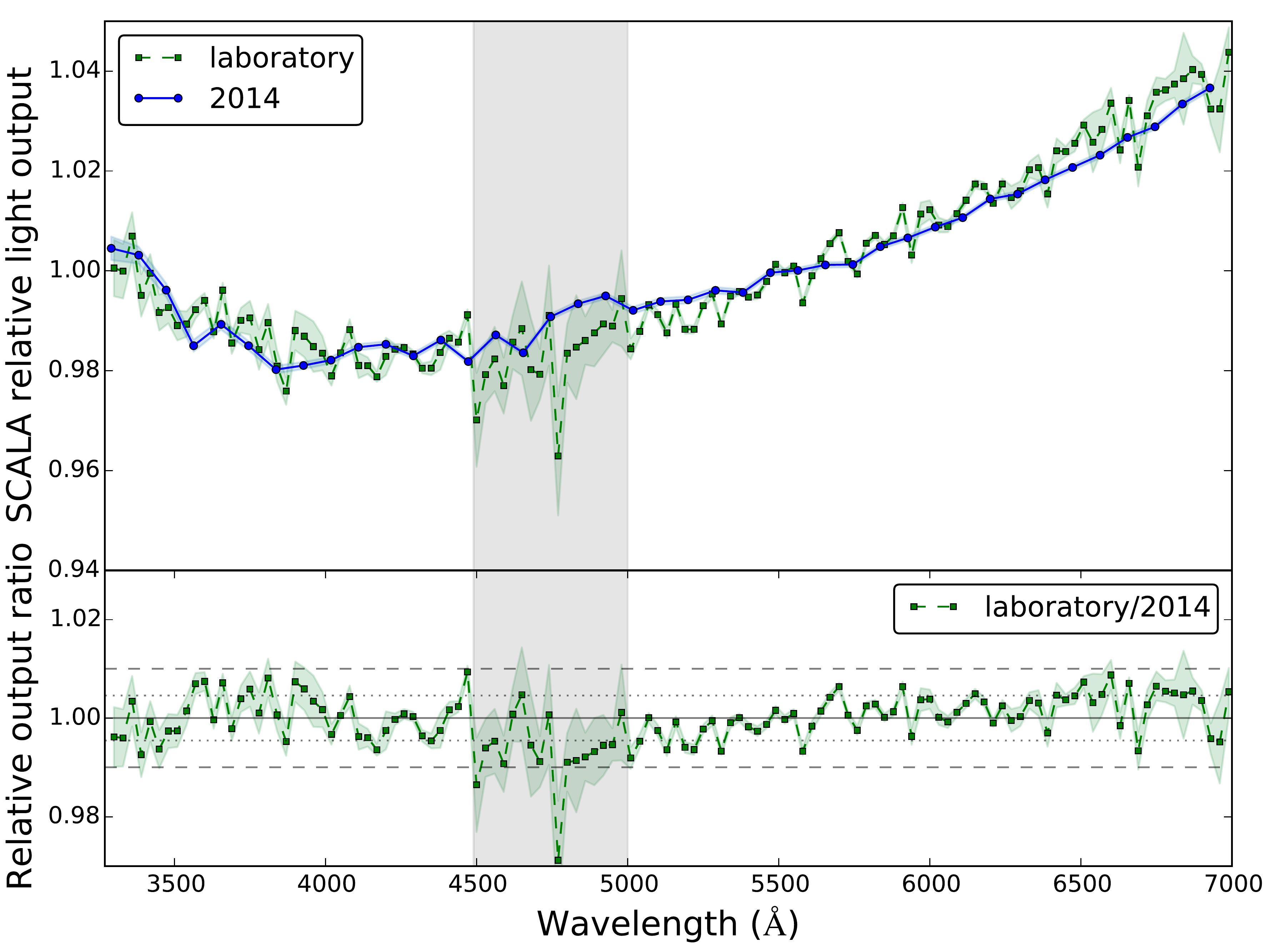}

\caption{The quantity $E_{s}$ is shown in the top panel. In green is the wavelength dependence of the SCALA relative light output estimated from the laboratory measurements.
In blue is that measured from the commissioning in May~2014.
Both have been normalized with respect to their average to compare their color trend. The bottom panel shows the ratio between the laboratory measurement and the 2014 commissioning measurements. The dashed lines represent the $\pm$1\% range around the averaged ratio of 1.000 (full line). The dotted lines are the measured standard deviation of 0.4\%. The vertical gray band delimits the region where weak emission lines of the Xe lamp are located (4500-5000\,{\AA}). The measurements have been performed with the Xe lamp and therefore they only go up to 7000\,{\AA}.}
\label{Flo:clap_ratio}
\end{figure}

By selecting the appropriate IS, fiber bundle and mirror and by combining these 18 beam response functions according to Equation~\ref{eq:module}, we can calculate the expected response of SCALA as fully integrated and operated at the telescope. 
The responses of ISs, mirrors and fiber bundles measured in the lab and plotted in Figure~\ref{Flo:bonn_test} are each scaled by the respective IS, mirror or fiber bundle response of the reference beam, thus mimicking the in-situ set-up wherein responses are measured relative to the reference CLAP. 
The sum over these 18 relative responses is shown with green squares in the top panel of Figure~\ref{Flo:clap_ratio}.
This curve is compared with the 2014 measurements of the fully integrated system, as determined using Equation~\ref{eq:a18} (blue circles). 
The measurements in this plot stop at 7000\,{\AA} since redward of this the laboratory tests used a Xe lamp whereas the in-situ data used the halogen lamp.
The measurements have been normalized to their average values since we are only interested in comparing their color trend and not their absolute values. 
These averages are computed over the wavelength range shown in the plot, and equal 19.12 for the lab measurements and 18.95 for the 2014 in situ measurements.
\begin{figure*}
\centering
\includegraphics[scale=0.26]{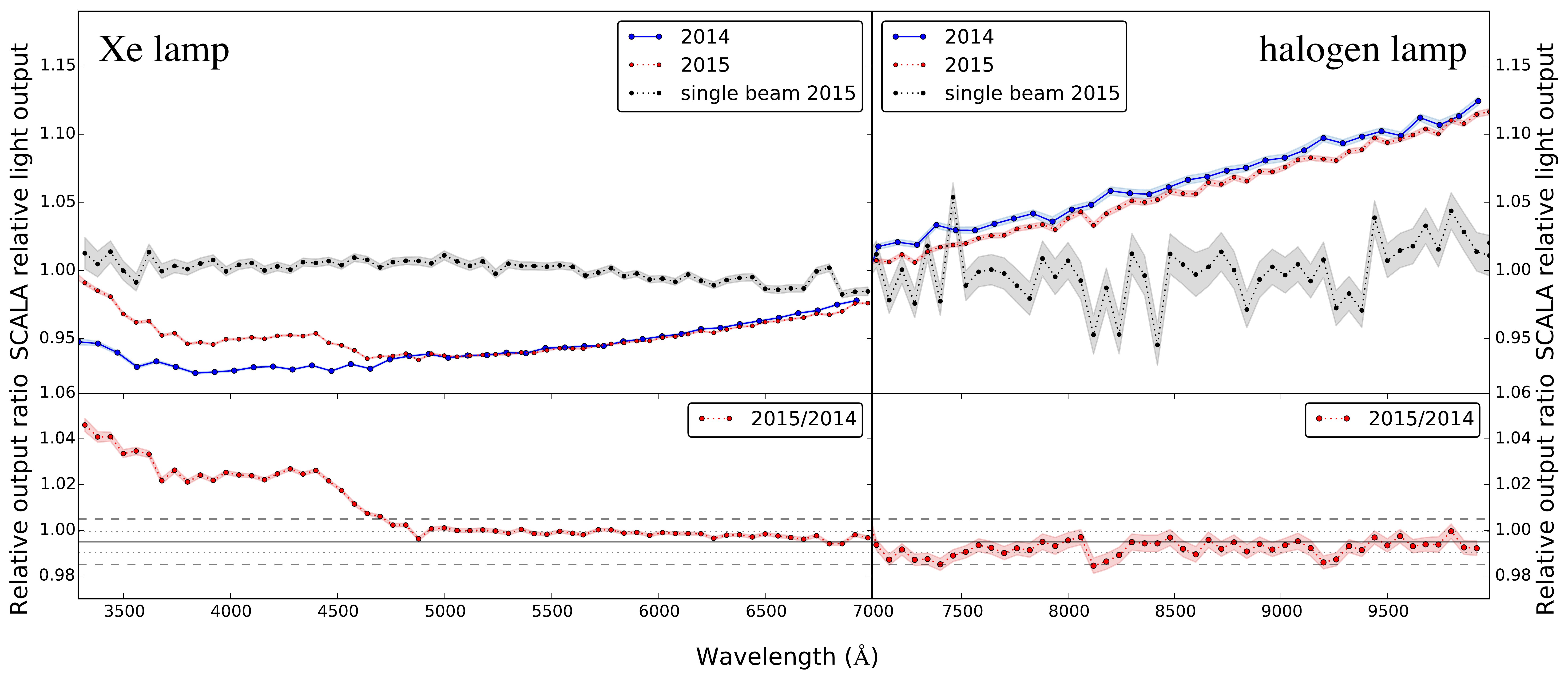}

\caption{The quantity $E_{s}$ is shown in the top panel for both figures, the SCALA relative light output from the commissioning in May~2014 is in blue and the latest commissioning in May~2015 is in red. Both are normalized to their average values to compare only their color trend. One of the mirrors illuminated by the same IS as the reference for the 2015 SCALA relative efficiency measurements, normalized by its average value, is shown using solid black circles. The bottom panels show the ratios between the two sets of commissioning measurements. The dashed lines represent the $\pm$1\% range around the averaged value computed for the measurements from 4700\,{\AA} to 10000\,{\AA}, where a gray offset is present. The full line, around 0.995, is the average and the dotted lines span the standard deviation interval ($\pm0.005$). The quantities in the left panel refer to $E_{s}$ measured with the Xe lamp while the right panel shows the measurements with the halogen lamp.}
\label{Flo:clap_ratio_hawaii}
\end{figure*}
As can be seen in Figure~\ref{Flo:clap_ratio}b, the ratio of the component-wise-combined measurements and the in-situ measurements in 2014 does not show a color trend over the wavelength range where the two sets of measurements. The scatter in the ratio is 0.4\%, which is within the expected errors (shown with a shadowed area around the line).
The color trend in the upper panel of Figure~\ref{Flo:clap_ratio}  is dominated by the relative transmissivity of the ISs, as shown in Figure~\ref{Flo:bonn_test} and discussed in Section~\ref{sec:control_is}.
Those two measurements were separated by only 2 months, but in that time the system was shipped and reassembled. That it stayed so consistent is very impressive and encouraging.
The agreement of the curves in Figure~\ref{Flo:clap_ratio} together with the consideration that they have been measured with independent setups, proves that SCALA is a highly reproducible system (at least for the overlapping wavelength range). It also demonstrates the success of our method, whereby the wavelength behavior of the SCALA relative light output is reconstructed based on photodiode measurements taken with only two of the SCALA beams at any one time. 
From this we conclude that for SCALA we understand the per-component causes of the wavelength behavior, and can reproducibly measure the system response.

\subsubsection{Commissioning measurements in 2014 vs 2015}
\label{sec:control2014-15}
By comparing in situ measurements taken over the year-long baseline between 2014 and 2015, we have detected a change in SCALA. 
Figure~\ref{Flo:clap_ratio_hawaii} shows $E_{s}$ for 2014 (blue circles) and 2015 (red circles). A significant difference can be seen in the bluer wavelengths in the 2015 estimation (left panels). 
These curves are normalized with respect to their average values of 20.09 and 19.46 for the 2014 and 2015 measurements respectively, computed over the full wavelength range.
The colored bands around the data points represent the statistical errors on the measurements.
They are larger on the right side of the plot due to the smaller statistics caused by the lower level of light produced by the halogen lamp.
The coarser wavelength sampling of the curves in 2014 relative to 2015 is due to a reduced fraction of time devoted to the  $E_{s}$ measurements.
In 2015 we opted for a more refined sampling and performed the measurements during three days.

One known difference is that the SCALA mirrors were cleaned before performing the measurements in 2015, and the reference mirror was cleaned more thoroughly compared to the others due to its easier accessibility.
From this alone, a relative difference between the different beams and the reference beam is expected.
Redward of 4700\,{\AA} the ratio between the responses for the two years is achromatic, and continues to be so beyond the lamp switch-over at 7020\,{\AA} (right panels of Figure~\ref{Flo:clap_ratio_hawaii}). 
Over this wavelength range the overall mean ratio is 0.995 (full line) and the standard deviation of the ratio is 0.005 (dotted lines).

This change, at the blue wavelength end, might be due to a more accentuated degradation of the reference beam response with respect to the other beams.
More specifically, a comparison between the responses of the two other beams belonging to the same IS as the reference beam do not show a deviation in the blue end, thereby excluding a relative degradation of the reference CLAP with respect to the other CLAP.
One of the beams illuminated by the same IS as the reference, normalized by its average value of 0.70, is plotted in Figure~\ref{Flo:clap_ratio_hawaii} with solid black circles and the label "single beam 2015". 
The fact that it is a smooth and mostly achromatic curve further suggests that the two CLAPs and the two mirrors did not degrade differently.
Instead, a color trend appears when the other beams from the other five ISs and fiber bundle arms are included in the comparison (red curve in Figure~\ref{Flo:clap_ratio_hawaii}).
This suggests that the change occurred in the IS and/or fiber bundle arm of the reference beam with respect to the others.

To explicitly test the reproducibility of our measurements, in 2015 we repeated the measurement of the relative responses of one of the beams on the first and  last days. 
The result of this test is discussed in Section~\ref{sec:scala_area_stb}, where we find our measurements to be reproducible at wavelengths above 4700\,{\AA}, but that a 0.8\% discrepancy exists blueward of this. This may be a partial contributor to the change seen between 2014 and 2015 for the bluest wavelengths shown in Figure~\ref{Flo:clap_ratio_hawaii}.

Together, these results point to excellent reproducibility in the  performance of SCALA redward of 4700\,{\AA}, and the need for further study of the behavior for the bluest 20\% of the wavelength range.

\section{"Telescope + SNIFS" calibration strategy}
\label{sec:strategy}

\begin{figure}[h]
\centering
\includegraphics[scale=0.23]{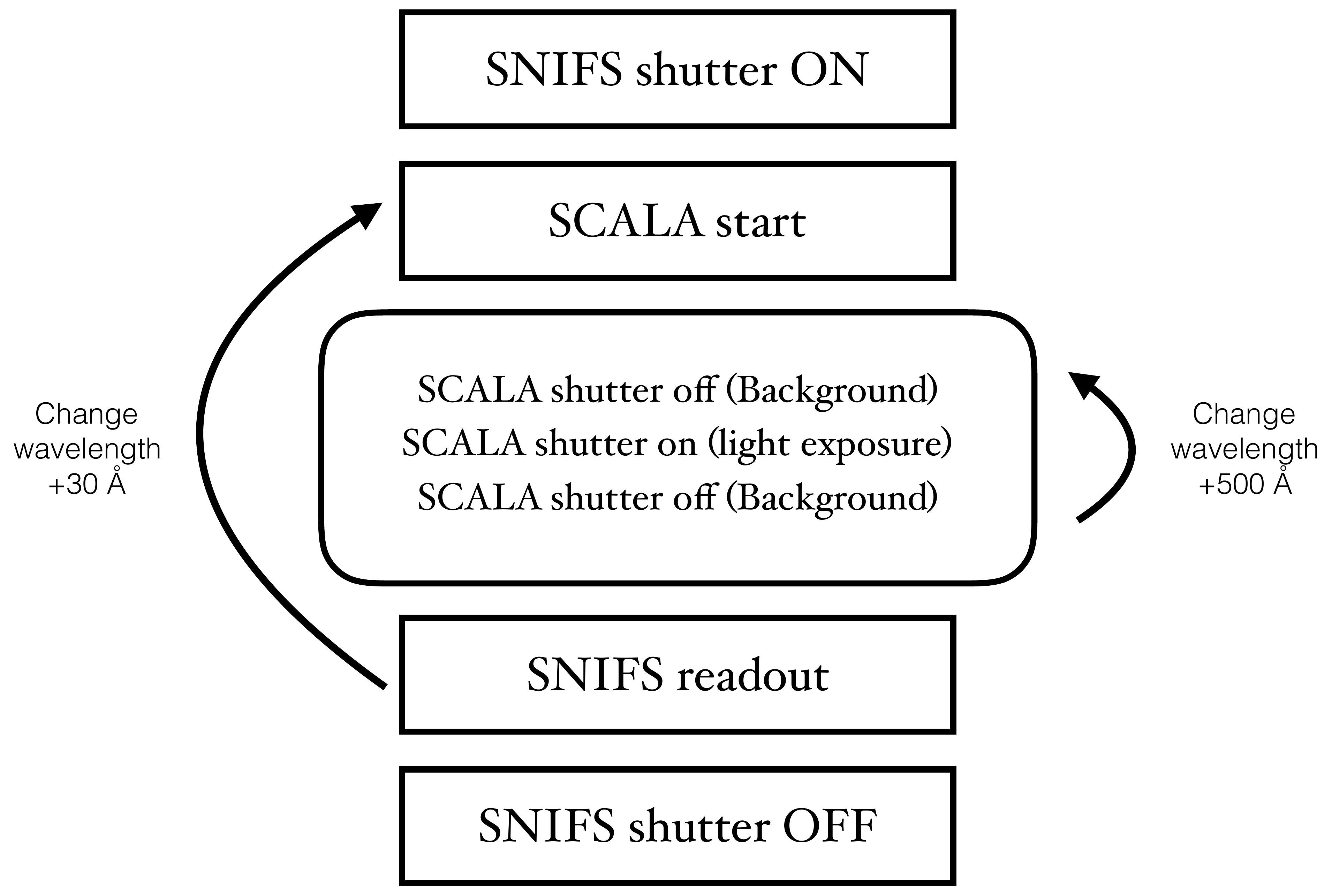}
\caption{Data taking scheme for a full wavelength calibration scan with SCALA. The rounded SCALA box represents the measurements performed by the CLAPs, while the SNIFS boxes represent observations with the spectrograph. The arrows indicate the loops used to generate observations of multiple SCALA wavelengths per SNIFS exposure, and to interleave these exposures to obtain dense sampling in wavelength over the full optical range.}
\label{Flo:data_take}
\end{figure}

\begin{figure}[h]
\centering
\includegraphics[scale=0.24]{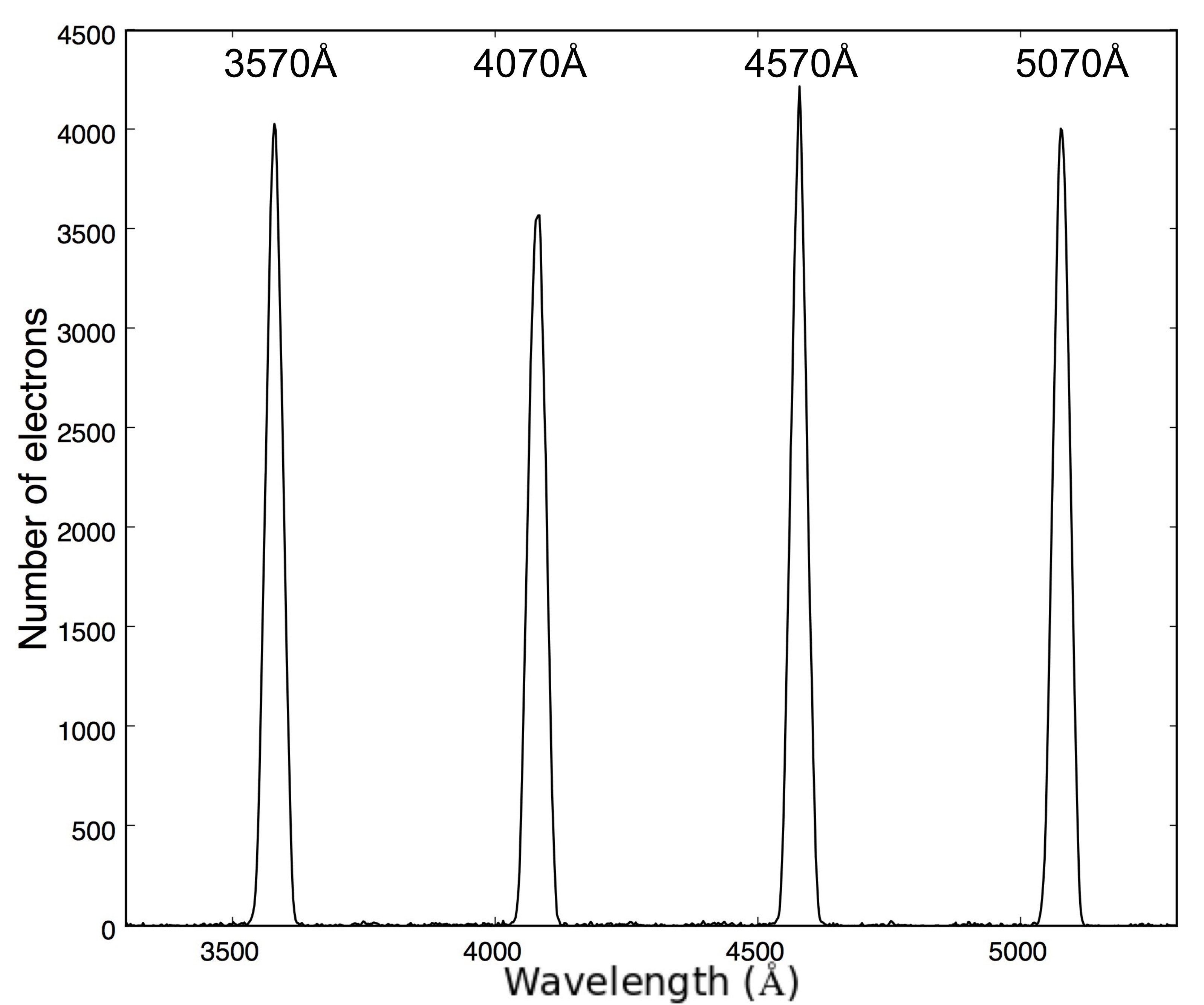}

\caption{Example of the wavelength-calibrated spectrum obtained from one of the spaxels of SNIFS for the blue channel. This is a typical exposure showing the 500\,{\AA} separation between the different wavelengths observed (the same as in Figure~\ref{Flo:clap_data}) and the nearly consistent number of electrons per wavelength due to a careful selection of the SCALA exposure time for each wavelength. The characteristic triangular shape and 35\,{\AA} FWHM of the SCALA line profile is also evident. }

\label{Flo:snifs_data}
\end{figure}

We now turn to the calibration of the telescope and instrument.
A schematic example of the calibration strategy is shown in Figure~\ref{Flo:data_take}.
In this conceptual diagram the block that shows background and SCALA light exposures refers to measurements by the calibrated photodiodes, as previously discussed in Section~\ref{sec:clap_datataking}.
To minimize time spent on readout of the SNIFS CCDs, we observe SCALA at several well-spaced wavelengths per SNIFS exposure, using the monochromator shutter to control the exposure time for each wavelength.
The SCALA line profile has a triangular shape with a FWHM of 35\,{\AA}. 
In K16 we verified that a separation of 500\,{\AA} between the wavelengths observed within one SNIFS exposure yields negligible cross-talk or overlap between adjacent wavelengths or adjacent spaxels.
Furthermore, we avoid crowded images or long exposures, by limiting the maximum number of wavelengths per exposure to four for the blue channel calibration (3300 - 5200\,{\AA}) and ten for the red (5000 - 10000\,{\AA}). A typical SNIFS exposure, after processing as in Section~\ref{sec:snifs}, is shown in Figure~\ref{Flo:snifs_data} for an observation of SCALA by SNIFS in the blue channel.
A SCALA observing sequence that calibrates the full SNIFS wavelengths range with a chosen wavelength sampling (from 30\,{\AA} on) therefore consists of multiple SNIFS exposures, each containing four or ten monochromatic lines, depending on the SNIFS channel.

The exposure time for each wavelength observation was set such that the corresponding $S/N>100$ per spaxel. The $S/N$ depends on both the number of photoelectrons from the combined SCALA output beams, the telescope throughput, and the sensitivity of SNIFS at each wavelength. The resulting exposure times range from 30\,s to 180\,s.
The overall time required for a complete ``$\mathrm{telescope}+ \mathrm{SNIFS}$'' calibration is 8\,hrs.
However, the flexibility of our acquisition software allows us to considerably reduce the exposure time for a calibration sequence, if desired.
One can, for example, use larger wavelength steps within each SNIFS exposures, which offers the possibility to calibrate SNIFS more quickly. For example, calibration in steps of 150\,{\AA} can be completed in less then 2 hours.
This option is very useful for test and commissioning runs, as well as for fast (and possibly daily) measurements of throughput.

\section{Pilot observing run}
\label{sec:data_take}

During the 2015 commissioning, from June~3 to 7, we performed a series of SCALA measurements with SNIFS during night-time, using the newly-commissioned aperture mask.
We alternated SCALA observations with standard star observations in order to check the stability of the ``$\mathrm{telescope}+ \mathrm{SNIFS}$'' calibration from SCALA during the night.
Due to non-photometric conditions and telescope issues, only the last night of observations, performed with an average seeing of 1.35~arcsec, was considered suitable for combining SCALA and standard star observations.

In the following subsections we detail our star/SCALA observation strategy for this night (Section~\ref{sec:obs_strategy}) and show the resulting throughput measurement (Section~\ref{sec:through}). Application of the SNIFS calibration to standard stars is on-going, so is not presented here.
\subsection{Observation strategy}
\label{sec:obs_strategy}
We subdivided the night into 5 sections.
We initially performed a $\sim2$\,hr SCALA calibration sequence, starting from 3330\,{\AA} with steps of 180\,{\AA}, during evening twilight, when the influence of ambient light in the dome is small compared to full daytime light. 
Then we opened the dome and performed standard star observations for $\sim2$\,hr. We closed the telescope dome and performed another SCALA calibration sequence with the same steps, but this time starting at 3390\,{\AA}.
We resumed observing stars until the end of the night, and finally took a SCALA calibration sequence in the morning starting at 3450\,{\AA}.
To ascertain the stability of the overall system during the night, e.g. SCALA alignment with the telescope and the ``$\mathrm{telescope}+ \mathrm{SNIFS}$'' throughput, a subset of nine wavelengths was reobserved during all three SCALA blocks during the night. We denote these monitoring sets of observations as A, B and C. 

Combining the SCALA calibration sequences performed during the night, provides us with 111 measurements of the instrumental response giving a complete SNIFS calibration sampled at 60\,{\AA} intervals.
The system stability can be examined by computing the ratios between the throughput of the reobserved wavelengths, A/B and C/B. From 4000 to 9000\,{\AA}, these two ratios are consistent with each other and centered around 1.001 with standard deviation of 0.003 (see Section~\ref{sec:scala_snifs_repro}). 
In the 3300-4000\,{\AA} region, the morning run (C) shows a 2\% offset.
Note that these are upper limits to the system stability as the A measurements were performed during evening twilight, the B measurements during night and the C measurements were performed after sunrise in the morning.
This results in different levels of ambient light in the dome, for which the software corrects using the CLAP background data segments, as shown in Figure~\ref{Flo:clap_data}.

\subsection{Throughput measurement}
\label{sec:through}
The throughput measurement is accomplished by comparing the observations of SCALA by SNIFS against those from the calibrated reference photodiode, according to the following equation:
\begin{equation}
{T(\lambda_{\mathrm{c}},\mathrm{spx})} = \frac{\int{I_{\mathrm{SNIFS}}(\lambda,\mathrm{spx})E'({\lambda})d\lambda}}{{C_{\mathrm{r}}(\lambda_{\mathrm{c}})}\cdot{D_\mathrm{r}(\lambda_{\mathrm{c}})}}\cdot{\frac{G}{E_{s}(\lambda_{\mathrm{c}})}}
\label{eq: trough}
\end{equation} 
where $I_{\mathrm{SNIFS}}(\lambda,\mathrm{spx})$ is the SCALA line intensity centered at $\lambda_{\mathrm{c}}$ observed by SNIFS in each spaxel, spx, in electrons and $E'(\lambda)$ is the energy of the photon for the wavelength $\lambda$; $C_{\mathrm{r}}$ is the integrated light measured by the reference CLAP (over the SCALA line exposure time); $D_\mathrm{r}$ is the factor to convert the reference CLAP measurement from ADU to W, as in Equation~\ref{eq:a18}; $E_{s}(\lambda_{\mathrm{c}}) $ is the SCALA light output relative to the reference CLAP from the 2015 measurements as in Equation~\ref{eq:a18} and shown in Figure~\ref{Flo:clap_ratio_hawaii} (without normalization this time);  $G$ is the geometrical factor that accounts for the different dimensions of the CLAP and SNIFS and also their different fields of view.
The geometrical factor, $G$, is determined by multiplying the ratio of areas of the CLAP ($5.8\times{5.8}$\,mm) versus area of a SCALA mirror against the ratio of the solid angles of the SCALA beam ($1^\circ$) versus that of a SNIFS spaxel.
Once scaled for the SCALA mirror and CLAP area, the factor $C_{\mathrm{r}}(\lambda_{\mathrm{c}})\cdot{D_\mathrm{r}(\lambda_{\mathrm{c}})}\cdot{E_{s}(\lambda_{\mathrm{c}})}$ provides the total illumination generated by SCALA as function of wavelength.

\begin{figure}[h]
\centering
\includegraphics[scale=0.245]{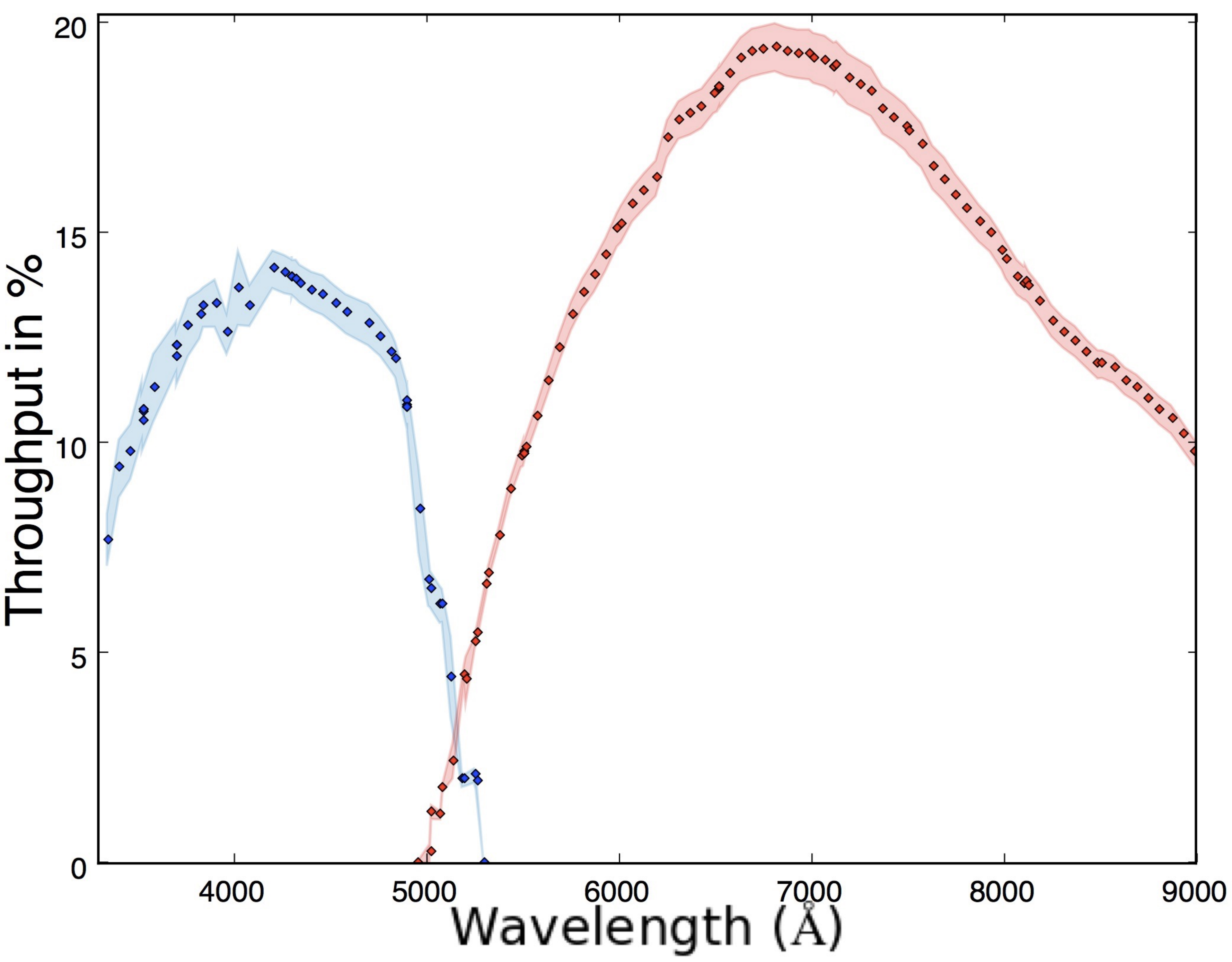}

\caption{The nighttime throughput, $T_{\lambda_i,spx}$, of the telescope plus SNIFS, averaged over all SNIFS spaxels. The blue and red circles represent the throughput measured by illuminating the blue and red spectroscopic channels, respectively, and the telescope with SCALA. The colored bands indicate the throughput standard deviation between SNIFS spaxels, and are thus \emph{not} an estimate of the throughput uncertainty.}
\label{Flo:through_scala0}
\end{figure}

From the average of Equation~\ref{eq: trough} over all spaxels we obtain the mean throughput curve shown in Figure~\ref{Flo:through_scala0}.
This curve has been measured by combining the three blocks of SCALA calibration sequences performed during the same night, as explained in Section~\ref{sec:obs_strategy}.

Application of this throughput measurement to the computation of the flux of a standard star is straightforward. The SNIFS optics reimage the telescope pupil onto the detector, thus, the fluxes from SCALA or from a standard star are extracted from the CCD detectors in an identical fashion when building a datacube. The datacube from SCALA is then used to flux-calibrate each wavelength of each microlens in the datacubes of standard stars; that is, it provides a surface-brightness calibration.
The final reported stellar flux for each wavelength results from a PSF fit over the $6.4"\times6.4"$ spatial extent of the microlens array. For typical seeing experienced with SNIFS on Mauna Kea, the microlens array contains more than 99\% of the light comprising the PSF model \citep{buton_thesis}. Therefore, the portion of the flux based on extrapolation of the PSF is quite small.

\section{Systematics}
\label{sec:systematics}

In this section we carefully examine the various sources of uncertainty. Some may be statistical, such as the repeatability blueward of 4700\,\AA, but we include them as systematics until more is known. The known or potential sources of systematic uncertainty are summarized in Table~\ref{tab: list_systematics}. Some of these errors are correlated because they are based on the same sets of calibration measurements and thus their statistical errors will be strongly correlated. Counting the statistical errors more than once would be incorrect. Thus, the net systematic uncertainty will be much less than the quadrature sum of the values listed in Table~\ref{tab: list_systematics}. The full covariance matrix, required for the final calibration, will be presented by K\"usters et~al. (in prep).
\begin{table*}[h]
\begin{center}
\caption{List of systematic uncertainties that contribute to the total uncertainty on the throughput measurements.}
\doublespacing
\begin{tabular}{lc}
\hline
Systematics & $\sigma_T/T$ \\ \hline
Ambient light in SCALA relative output (daytime clouds) & $<0.5$\%  \\
Ambient light in SCALA relative output (daytime clear) & $<0.4$\%  \\
Ambient light in nighttime CLAP data & $<0.2$\% \\
Wavelength uncertainty on SCALA relative output & $<0.021$\% \\
SCALA relative output stability & \begin{tabular}{c}
$<0.7$\%\,${\lambda}<4700$\,{\AA}\\[-0.5em]
$<0.02$\%\,${\lambda}\geq 4700$\,{\AA}\end{tabular}\\
Inhomogeneous color illumination of the entrance pupil & $<0.09$\% \\
Stray light in $0.5^\circ$ beam & $<0.51$\% \\
CLAP integration over time error & $<0.07$\% \\ 
Optical cross talk in SCALA & $<0.05$\% \\
Optical cross talk in SNIFS & $<0.1$\% \\
SCALA and SNIFS reproducibility & \begin{tabular}{c}
0.3\%\, for ${\lambda}\geq$4000\,{\AA}\\ [-0.5em] 
2\%\, for ${\lambda}$<4000\,{\AA}\end{tabular}\\
CLAP calibration & \begin{tabular}{c}
$0.74$\%\, for ${\lambda}<3400$\,{\AA}\\[-0.5em]
$0.32$\%\, for $3400 < {\lambda} < 4700$\,{\AA}\\[-0.5em]
$0.18$\%\, for ${\lambda}\geq 4700$\,{\AA}\end{tabular}\\
\hline
\end{tabular}
\label{tab: list_systematics}
\end{center}
\end{table*} 

Of the measurements composing Equation~\ref{eq: trough}, the biggest contribution to the uncertainty is due to the $E_{s}$.
This particular set of measurements has been performed under rather different conditions than the observations of SCALA by SNIFS, making it subject to several sources of contamination:
\begin{itemize}
\item since these measurements were made during daytime, they can have residual ambient light contamination (see Section~\ref{sec:ambient_ligh}),
\item since these measurements do not have associated SNIFS observations to provide precise wavelength measurements, we need to consider the uncertainty caused by the limit reproducibility of the monochromator wavelength setting (see Section~\ref{sec:wave_uncertain}),
\item reproducibility of the photodiode position for each mirror when cross-calibrating each beam (Section~\ref{sec:scala_area_stb}).

\end{itemize}

In the following we start with these systematics, both for the nighttime and daytime measurements, and then consider the systematics related to the other elements of Equation~\ref{eq: trough}. 
We also note that $E_{s}$ measurements have higher statistical uncertainties due to their shorter exposure times.

\subsection{Ambient light contamination}
\label{sec:ambient_ligh}
The throughput measurements derived using Equation \ref{eq: trough} were performed mostly during nighttime with the telescope dome closed. 
Ambient light contamination can, therefore, be neglected in the analysis of the calibrated photodiode data.
On the other hand, $E_{s}$ was measured during daytime and, therefore, dome light leaks could lead to varying levels of ambient light contamination.

During the day there are two cases that can occur: diffuse and scattered ambient light that smoothly drifts with time, and scattered ambient light that quickly varies on a time scale of seconds, e.g. due to clouds crossing the sky.
Variations on timescales longer than individual SCALA line exposures can be accurately accounted for by interpolating the CLAP background measurements.
For these cases the background measurements before and after the light exposure (Figure~\ref{Flo:clap_data}) are representative of the dark current and ambient light level during the light exposure. 

The partly-cloudy daytime case is more complicated.
To estimate the uncertainty on background removal we have obtained a series of daytime (and nighttime) CLAP measurements with only ambient light, as shown in K16.
We tested the precision of the background reconstruction by artificially dividing these data into background and signal segments, and measuring how well we retrieve the background.
From such tests we find that a simple linear interpolation of the background shows good results, with a reconstruction of high-frequency background variations that is better than 6\% of the faintest (longest exposure) line emitted by SCALA. However, as the $E_{s}$ determination relies on ratios between simultaneous measurements from two different CLAPs and the background misestimates are strongly correlated, the error on the signal ratio reconstruction is mostly compensated. The net error is of order of 0.5\%.

Datasets for only a few mirrors exhibit such fast variations during periods when $E_{s_{\lambda_i}}$ was being measured in 2015.

In K16 we showed that for diffuse and scattered ambient light that smoothly drifts with time, the background misestimate of the SCALA relative light output is less than 0.4\% of the faintest line emitted by SCALA.
As already mentioned, ambient light is not a concern during nighttime measurements, where CLAP dark current variations can be reconstructed to better than 0.2\% of the faintest (longest exposure) line emitted by SCALA.

A possible way to further suppress this contamination, whether measurements are made during the day or night, would be to perform the SCALA wavelengths observations with additional background samples introduced during the light exposure measurements.
That is, closing the monochromator shutter during the light exposure would give a better sampling of possible changes in ambient light, especially for long exposures. This would not increase the exposure times needed by SNIFS by much.
By increasing the sampling frequency of the photodiode data, a constant statistical uncertainty could be maintained, as the error would be computed on a larger sample.

\subsection{Wavelength uncertainty}
\label{sec:wave_uncertain}
Our calibration device produces monochromatic light through a monochromator, for which the stated wavelength reproducibility is about 1\,{\AA} according to the manufacturer.
For the SNIFS system throughput estimation we use the wavelength calibration from SNIFS data to evaluate the central wavelengths observed, therefore this error does not directly appear in the systematics budget.
It is present, though, in the determination of SCALA relative light output,  $E_{s}$, since in this case, to save time, we do not obtain any associated SNIFS exposures.

\begin{figure}[h]
\centering
\includegraphics[scale=0.40]{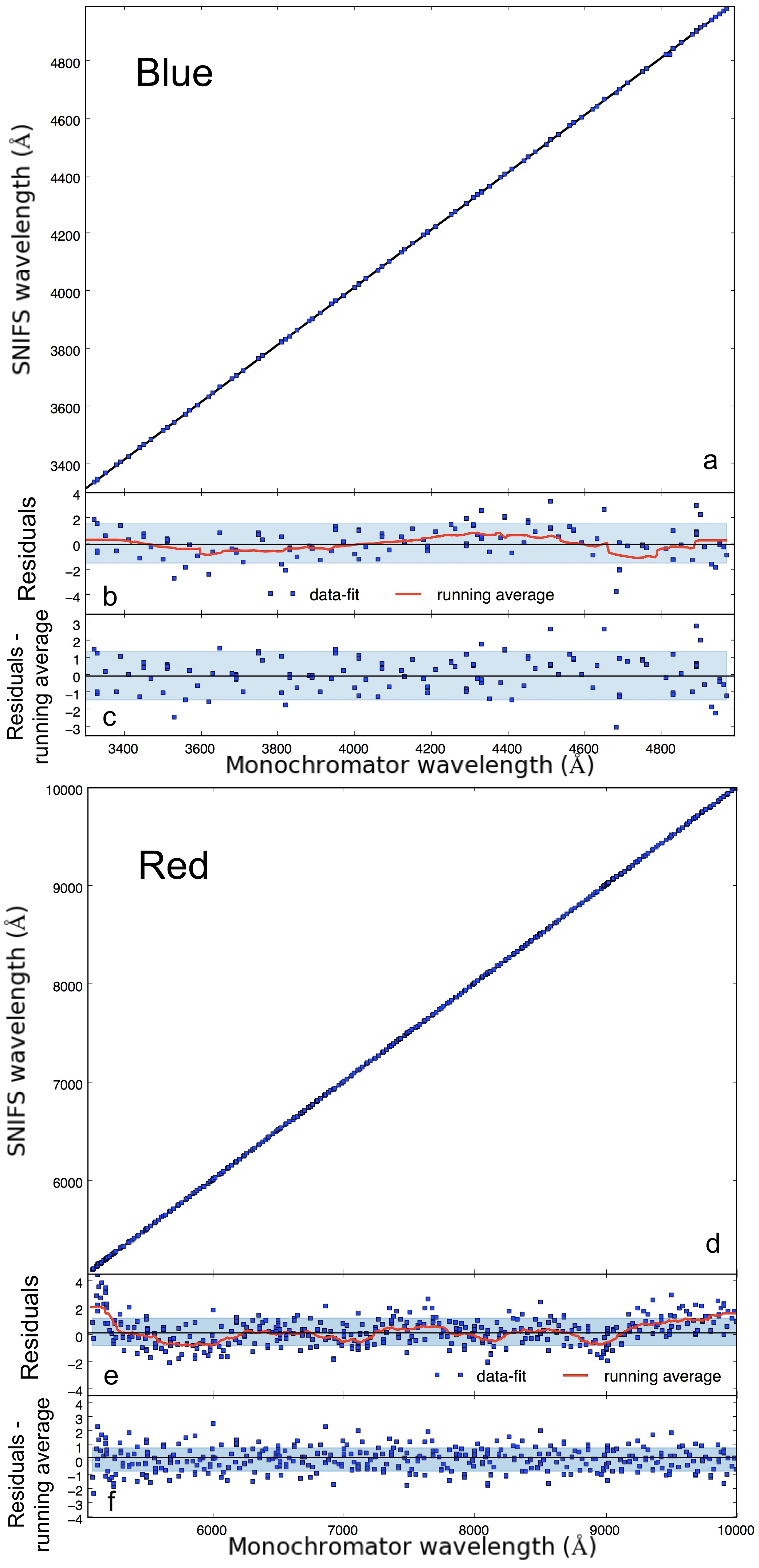}

\caption{Wavelengths measured using the SNIFS exposures for the blue channel (panel a) and the red channel (panel d) versus the ones requested to the monochromator, plotted together with the linear fit (black line). Panels b and e show the residuals from the linear fits (blue squares) and running average of the residuals (red line), computed with an averaging window of 20 data points. The blue shaded region represents the standard deviation interval of the data of $\pm$1.51\,\AA\ (in panel b) and $\pm$1.01\,\AA\ (in panel e). Panels c and f show the results of subtracting the running averages from the residuals of the linear fits (blue squares). The blue shaded region represents the standard deviation interval of the data of $\pm$1.39\,\AA\ (in panel c) and $\pm$0.79\,\AA\ (in panel f).}

\label{Flo:wave_reprod}
\end{figure}

We have used the SNIFS wavelengths calibration to verify the stated wavelength precision by comparing the central SCALA wavelengths measured from a set of SNIFS exposures with those requested of the monochromator and recorded in the headers of the CLAP photodiode data.
These follow a linear trend, which can be fit for the blue and red wavelengths as shown in Figure~\ref{Flo:wave_reprod} (panels \ref{Flo:wave_reprod}a and \ref{Flo:wave_reprod}d) together with their residuals (panels \ref{Flo:wave_reprod}b and \ref{Flo:wave_reprod}e).
We find a residual trend after subtracting the linear fit to the data,
with standard deviations of 1.51\,\AA\ and 1.01\,\AA\ for the blue and red channels, respectively. Under the assumption that the deviations are due to a stable mechanical effect, monochromator wavelength errors can be decreased further by subtracting the 20-point running average of the residuals. The standard deviation of these residuals is 1.39\,{\AA}, for the blue channel (Figure~\ref{Flo:wave_reprod}c), and 0.79\,{\AA}, for the red channel (Figure~\ref{Flo:wave_reprod}f). These reductions in the standard deviations have high statistical significance, so are not simply the results of random scatter.
This combination of linear fits and running averages for the spectroscopic channel is used to fine-tune the monochromator wavelength in the determination of $E_{s}$.

Such small errors in the wavelength estimates do not constitute an issue for the $E_{s}$ measurement. When recomputing $E_{s}$ using a wavelength that is artificially increased (or decreased) by one standard deviation, we find the same result within 0.021\% (in the worst case).

Finally, we have compared multiple observations when the same wavelength was requested to the monochromator. These are consistent with the monochromator wavelength reproducibility stated by the manufacturer. 
  
\subsection{SCALA relative light output stability}
\label{sec:scala_area_stb}

\begin{figure}[h]
\centering
\includegraphics[scale=0.245]{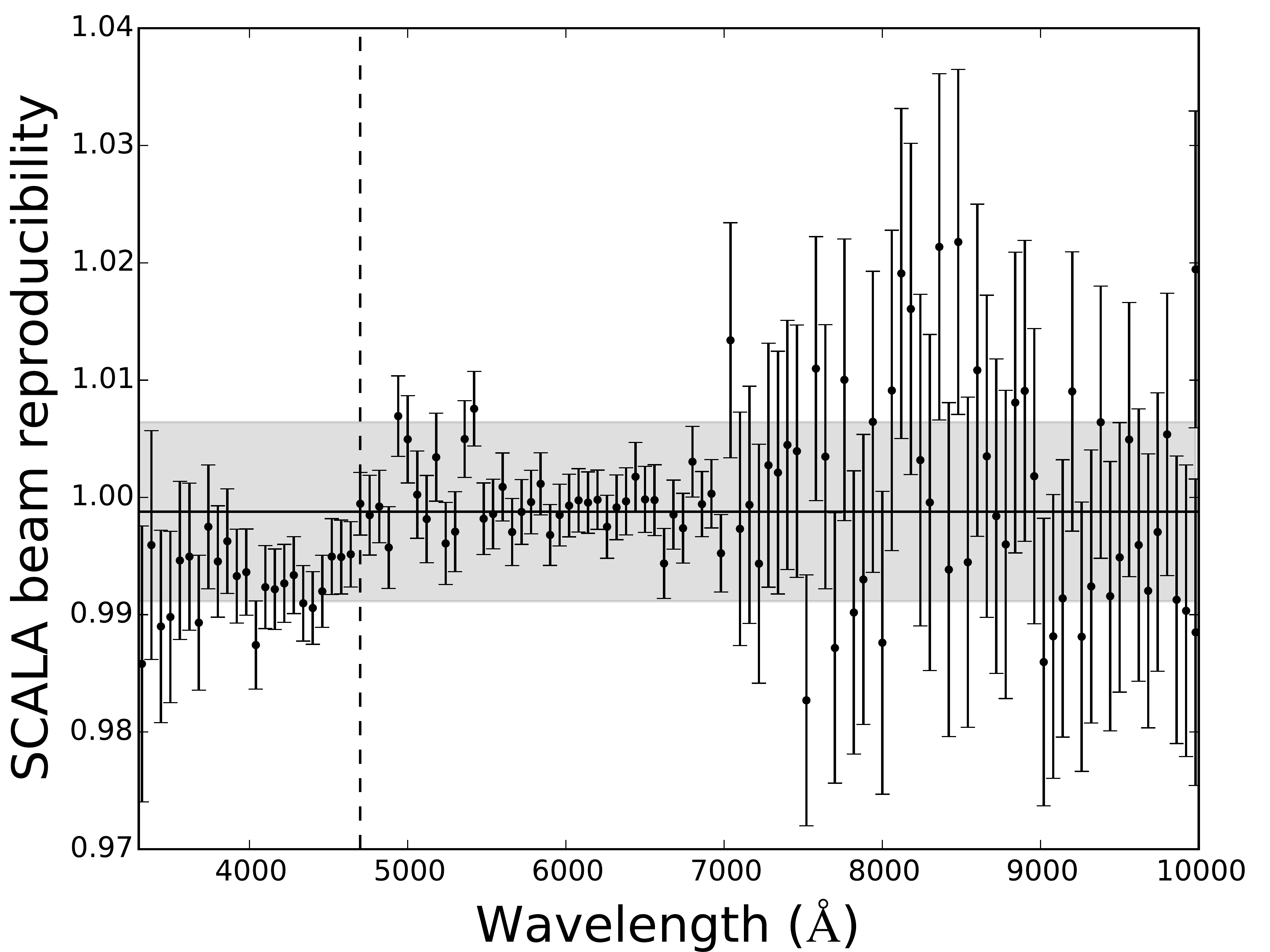}
\caption{The ratio, $\alpha/{\beta}$, between repeated measurements of the same SCALA beam is shown as black circles. The black line represents the average of $0.999$ and the shaded gray region brackets the standard deviation interval of $\pm$0.8\%. The vertical dashed line separates the data above and below 4700\,{\AA}. Redward of this the repeated measurements are highly reproducible, but blueward of this there are clear variations.}
\label{Flo:reprod}
\end{figure}

We have tested the reproducibility of the SCALA relative light output, $E_{s}$, by repeating the measurement of one of the 18 SCALA beams at the start and end of the four day calibration run.
We denote these sets of CLAP ratios between the moving and reference photodiodes as $\alpha$ and $\beta$, respectively.
The mean reproducibility is excellent, having a mean of 0.999 (blue line in Figure~\ref{Flo:reprod}) and a standard deviation of 0.008.
The measurements at the reddest wavelength range are noisier only because they have lower flux levels. 
These fluctuations  are further suppressed when all 17 beams are combined.
In regular nighttime observation we expose longer for wavelengths with lower intensity emission, and this maintains a constant $S/N$.
When measuring the SCALA relative light output, such long exposure times are not feasible as the measurement has to be made 18 times (once for each non-reference beam + the reproducibility sequence). For this test we used a constant exposure time of 5\,s per wavelength.

As can be seen in Figure~\ref{Flo:reprod}, the ratio between sets $\alpha$ and $\beta$ shows a 0.8\% difference for blue wavelengths ($<4700$\,{\AA}), likely due to a change in the system on time scales of a few days.
Relative evolution of the IS and/or fiber bundle arm belonging to the reference projector module with respect to the other SCALA modules, as mentioned in Section~\ref{sec:control2014-15}, is a possible cause.
Measurements of the beams from the same reference module do not show this evolution, as visible from the plot of one of them in Figure~\ref{Flo:clap_ratio_hawaii}.
Note that SCALA relative light output measurements were performed in a mixed order -- not module by module -- and the reference module beams were measured during the second day. 
To compute the contribution to the systematic error, we assume that this change is bracketed by the ratio of $\alpha$ to $\beta$, as $\beta$ was the final measurement made and the nighttime measurements (for which we use the SCALA light output) were acquired just before the $\alpha$ data-set.

This systematic error is propagated to each of the SCALA beam measurements, except for the two beams that belong to the reference module and thus are illuminated by the same fiber bundle arm and IS. As each beam ratio has a similar amplitude, the total error on the SCALA relative light output is of the same order as the one from a single beam, $<0.7$\%.

\subsection{Inhomogeneous illumination of the entrance pupil as a function of wavelength}
\label{sec:inhomo}
\begin{figure}[h]
\centering
\includegraphics[scale=0.24]{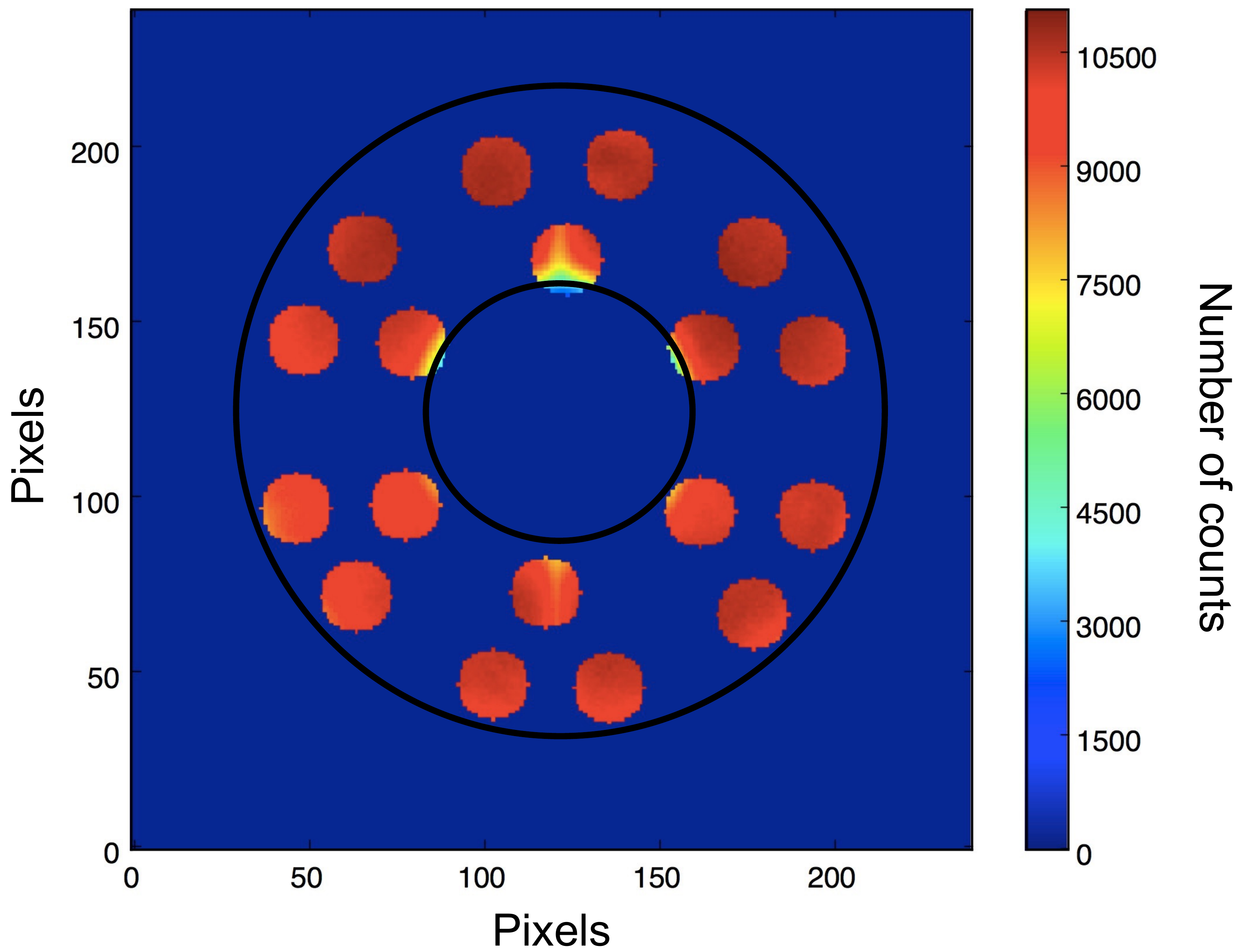}
\caption{Entrance pupil image of twilight, from the SNIFS imaging channel with pinhole array. The image has been masked to mimic the SCALA illumination pattern. The black circles represent the entrance pupil of the telescope with the secondary mirror obscuration.}
\label{Flo:pupil}
\end{figure}

As the SCALA beams have different relative color efficiencies, they will weight the reflectivity of the corresponding regions of the primary mirror differently. 
To estimate the potential size of such an effect might have, we use twilight sky images taken with the pinhole array in the SNIFS imaging channel. Many such images are stacked together to create an entrance pupil image free of CCD spatial features. Since these images are from nights without the SCALA pupil mask mounted, we mask the full-aperture pupil image in such a way so as to reproduce the SCALA illumination pattern. The result is shown in (Figure~\ref{Flo:pupil}).
To calculate the beam inhomogeneity effect, we first scale each of the regions in the pupil illuminated by SCALA with the corresponding SCALA beam relative light output (measured in 2015) binned every 300\,{\AA}, and then sum over the full pupil image.
For comparison, we sum over the full pupil image (i.e., without any scaling applied), and then scale the result by the total SCALA relative light output ($E_{s}$, as measured in 2015), also binned every 300\,{\AA}.
This mimics a real analysis, in which we do not know how the reflectivity of the primary mirror weights the SCALA beams.
The ratio from these two approaches provides the systematic error due to the non-uniform illumination of the entrance pupil of the telescope generated by SCALA, as a function of wavelength.
We find that these two weightings have differences $<0.09$\%.
The pinhole array is mounted in front of a blue filter, so we do not have the entrance pupil illumination over the entire wavelength range used here.
However we can assume that 0.09\% is an upper limit on this systematic error as the mirror reflectivity appears less homogeneous for bluer wavelengths.

\subsection{Stray light}
\label{sec:straylight}

The well-collimated beams of SCALA are designed to minimize the amount of scattered light.
However, the beam aperture is still $1^\circ$, so rays far off axis might be a source of stray light within the telescope or SNIFS.
To study this question we mounted smaller apertures in the IS output ports, thus reducing the SCALA angular extent to half its original size.

Using SCALA to illuminate the SNIFS photometric channel with white light, we produced several flat field images in $u, g, r, i, z$ filters and compared the results between the usage of the reduced $0.5^\circ$ beam and the regular $1^\circ$ beam.
As shown in K16, these have smooth spatial gradients across the entire SNIFS imaging channel, and there is no wavelength dependence. We also observed a set of monochromatic SCALA lines with both $0.5^\circ$  and $1^\circ$ SCALA beam width using the spectroscopic channels of SNIFS. 

\begin{figure}[h]
\centering
\includegraphics[scale=0.24]{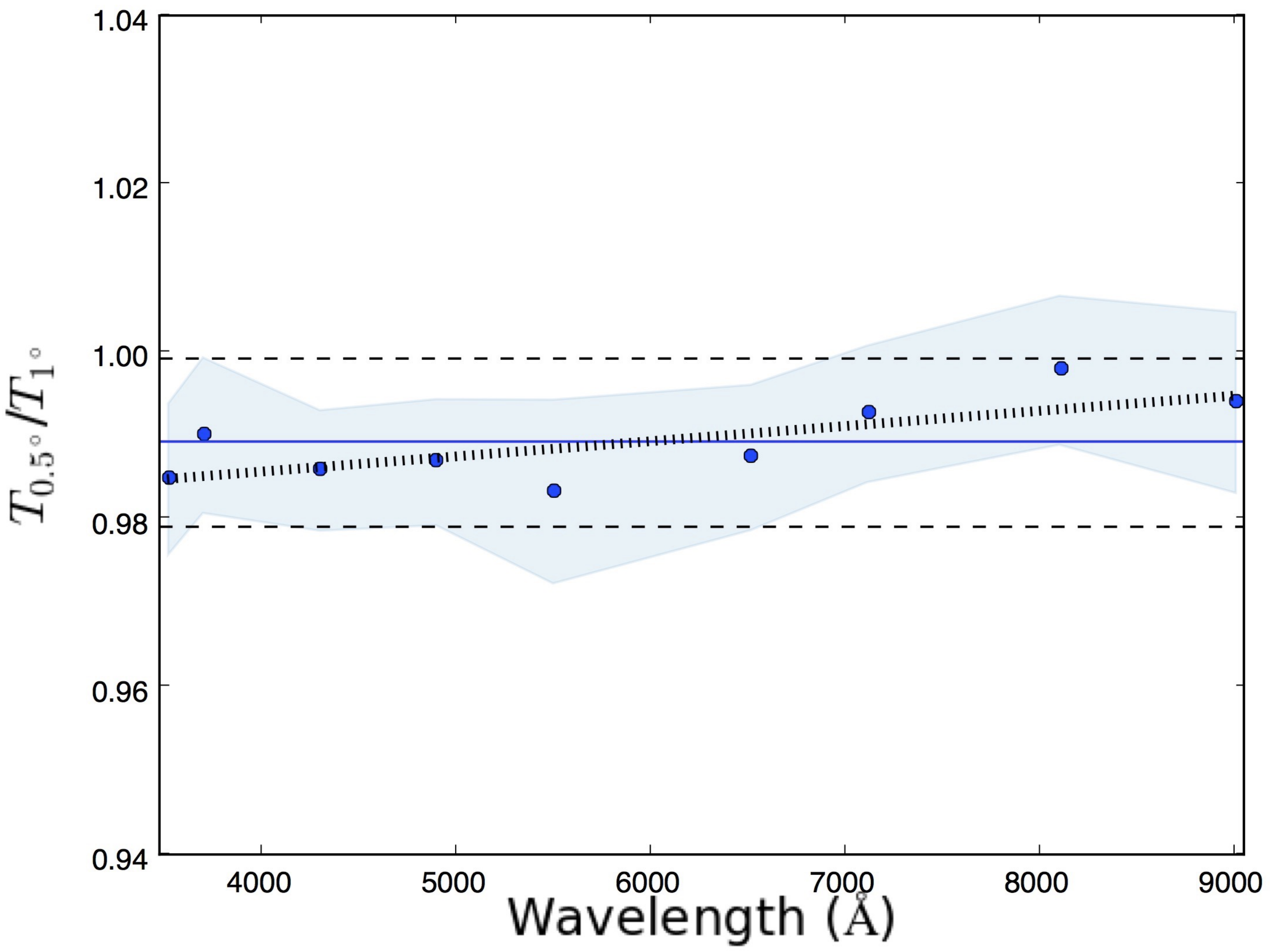}

\caption{The ratio between the quantity $T_{\lambda_i}$ is shown for the same SCALA wavelengths observed with $0.5^\circ$  and $1^\circ$ SCALA beam widths. An offset of 0.989 (blue full line) is present. There is no strong trend with wavelength. The values across wavelengths agree to much better than 1\% (black dashed lines), having a standard deviation of 0.4\%. The shaded area around the circles is an indication of how much the ratio varies from spaxel to spaxel. The dotted line is a linear fit to the data showing that a small slope may remain (see text)}.

\label{Flo:stray_light}
\end{figure}

For the analysis of these data it is necessary to consider the fact that the alignment of the telescope with SCALA was performed using the full $1^\circ$ beam. Investigation showed that some of the beams formed using the reduced IS apertures were not sufficiently well aligned to give full illumination to SNIFS.
This misalignment was found by visually inspecting the entrance pupil images acquired by the SNIFS imaging channel with the pinhole array. These images are acquired by default at the same time of observations of SCALA with the spectroscopic channels of SNIFS.
We found that of the $18$ beams, one was not visible at all when using the smaller IS aperture, while another was only partially visible.
This suggests that one of the mirrors was misaligned in such a way that it was not able to illuminate the focal plane of the telescope, and another one was providing only partial illumination.

While this misalignment does not constitute an issue for the regular SCALA observations performed without beam reducer, with all $1^\circ$ SCALA beams illuminating the focal plane, the small-aperture light levels need to be corrected for the missing light in order to meaningfully compare the two sets of measurements. This was done by
summing the relative efficiencies of only the $16$ well-aligned SCALA beams and half of the relative efficiency of the partially visible beam when calculating $E_{s}$ using Equation~3 for the $0.5^\circ$ beam. 
The ratios between the $0.5^\circ$ and $1^\circ$ SCALA beam-width measurements are shown in Figure~\ref{Flo:stray_light}. 
As detailed in K16, any trend with wavelength is below 1\%. 
A linear fit from blue to red wavelengths has range of $\pm 0.51$\% over the wavelength range of interest, with a borderline significance of 2.9$\,\sigma$.
Including a greater or lesser fraction of the partially illuminating beam does not increase the slope of this linear fit.

We thus find no signs of significant wavelength-dependent stray light. We are currently planning future tests of stray light using even smaller IS port apertures and correspondingly narrower output beam widths.

\subsection{Exposure time error}

In order to minimize the error on the SCALA exposure time our method is to integrate over the entire CLAP exposure data using Simpson's method.
The sampling frequency of 1~kHz allows us to sample up and down the ramp of the signal due to the monochromator shutter opening and closing.
With this sampling, the light emitted in the up and down ramps amounts to less than 0.07\% of the total light measured during an exposure.
Additionally, by design the shutter does not produce a differential illumination of CLAP and SNIFS when opening and closing.

\subsection{Optical cross talk within SCALA and SNIFS}

There are two possible sources of optical cross talk for our calibration observations.
First the CLAP photodiode might receive scattered light from other beams.
Second, in the SNIFS datacube, part of the light belonging to one spaxel might be collected in another spaxel. In the latter case, because the location of each spaxel illuminated with a given wavelength is staggered with respect to the CCD rows, such light would be displayed at other wavelengths.
In K16 we showed that this second effect is negligible, since the light from another spaxel is <0.1\% of the light integrated in the monochromatic line.

The cross talk due to scattered light from another SCALA beam hitting the calibrated photodiode was measured in K16 by closing the integrating sphere shutter belonging to the reference SCALA beam. The amount of light observed by this indirect illumination of the CLAP photodiode is achromatic up to 7000\,{\AA}, with a relative contribution of 2.5\%, and then linearly increases to a relative contribution of 5.7\% by 9700\,{\AA} (Figure~4 in K16).
The measurement error is <0.05\%.
This effect is mostly due to the reflection properties of the black anodized aluminum SCALA support structure and therefore should be constant over time, as repetition of this test confirmed.
The reference calibrated photodiode measurements, $C_{\mathrm{r}}$ in Equation~\ref{eq: trough}, have been corrected for this indirect illumination. 
In the SCALA relative light output measurements given by Equation~\ref{eq:a18}, the ratio between the two CLAP photodiode measurements highly suppresses the cross talk effect, since the different beams have similar amounts of contamination.

\subsection{Scattered light from monochromator}

The monochromator out-of-bandpass light leakage quoted by the manufacturer is 0.15\%.
This is further reduced by the SNIFS background subtraction software, yielding a final 0.05\% contamination from an entire line. 
This is a lower limit evaluated from the background level between SCALA lines in the SNIFS datacube.
However the CLAP photodiode integrates the entire monochromator spectrum, weighting it with its own responsivity. Therefore, even such a low level of scattered light could cause a bias as large as 1\%.
This is currently the limiting systematic uncertainty with SCALA. We are now performing narrow filter measurements to quantify this effect. If we confirm a 1\% effect, a second monochromator in series with the first could be installed to effectively remove this contamination.

\subsection{SCALA and telescope+SNIFS reproducibility}
\label{sec:scala_snifs_repro}
\begin{figure}[h]
\centering
\includegraphics[scale=0.25]{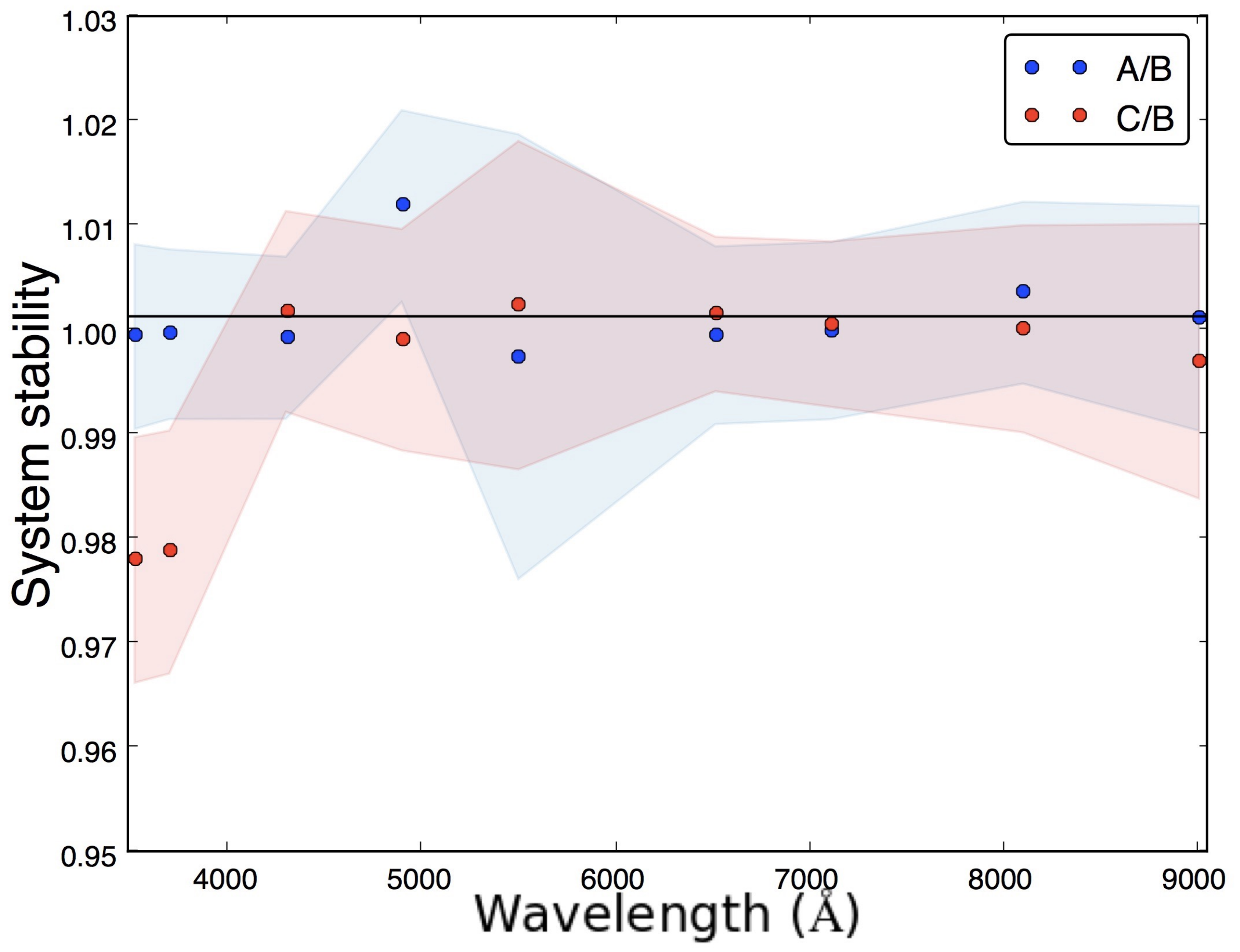}

\caption{Ratios of the throughput measured from the same SCALA lines observed during the three (A, B, and C) calibration sequences taken June~7, 2015. The shaded blue region indicates the spaxel to spaxel variations. The black line is the average of the ratios from 4000\,{\AA} onwards, which is $1.001$.}

\label{Flo:same_wave}
\end{figure}

As mentioned in Section~\ref{sec:data_take}, for each of the three SCALA observing block sequences during June~7, 2015, we observed a set of lines in common.
The system stability can be examined by computing the ratios, A/B and C/B, between the throughput of the reobserved wavelengths. 
From 4000 to 9000\,{\AA}, the two ratios are consistent with each other and centered around 1.001 with standard deviation of 0.3\% (Figure~\ref{Flo:same_wave}). 
In the 3300-4000\,{\AA} region, the morning run (C) shows a 2\% offset.
The origin of this difference is unknown. Repeated observations of this wavelength region may be able to reduce the net calibration error.
But for now we take the full amplitude of the offset as a systematic uncertainty. 
Above 4000\,{\AA} the reproducibility of our system is very good and is not a limitation.
 
\subsection{CLAP photodiodes}

Of course our calibration accuracy can only be as good as that of the reference system we are using. The CLAP photodiodes were calibrated by the DICE team \citep{Dice} and have uncertainties -- mostly dominated by the NIST photodiode calibration uncertainties -- around 0.74\% between 3200 and 3400\,{\AA}, then $<$0.32\% up to 4700\,{\AA} and $<$0.18\% between  4700 and 9500\,{\AA}.
Here we have assumed that the 2$\sigma$ uncertainties quoted by NIST are Gaussian, so we can convert them into 1$\sigma$ uncertainties. This error would be statistical if we used many different CLAPs, but since we primarily use the reference CLAP, we take this as a systematic uncertainty.

\section{Other potential applications}
\label{sec:potential_use}
There are several possible additional applications for SCALA, some of which are currently being applied to SNIFS:
\begin{enumerate}
\item Directly quantify the zero- and second-order light in the spectrograph for estimating whether it contaminates the first-order light.
\item Stray light study in the telescope optics, as briefly shown in the Section~\ref{sec:straylight}.
\item Individual SCALA wavelength exposures can be used to measure cross-talk between spaxels.
\item Filter+CCD curves for the SNIFS imaging channel filter set.
\end{enumerate}

\begin{figure}[h]
\centering
\includegraphics[scale=0.245]{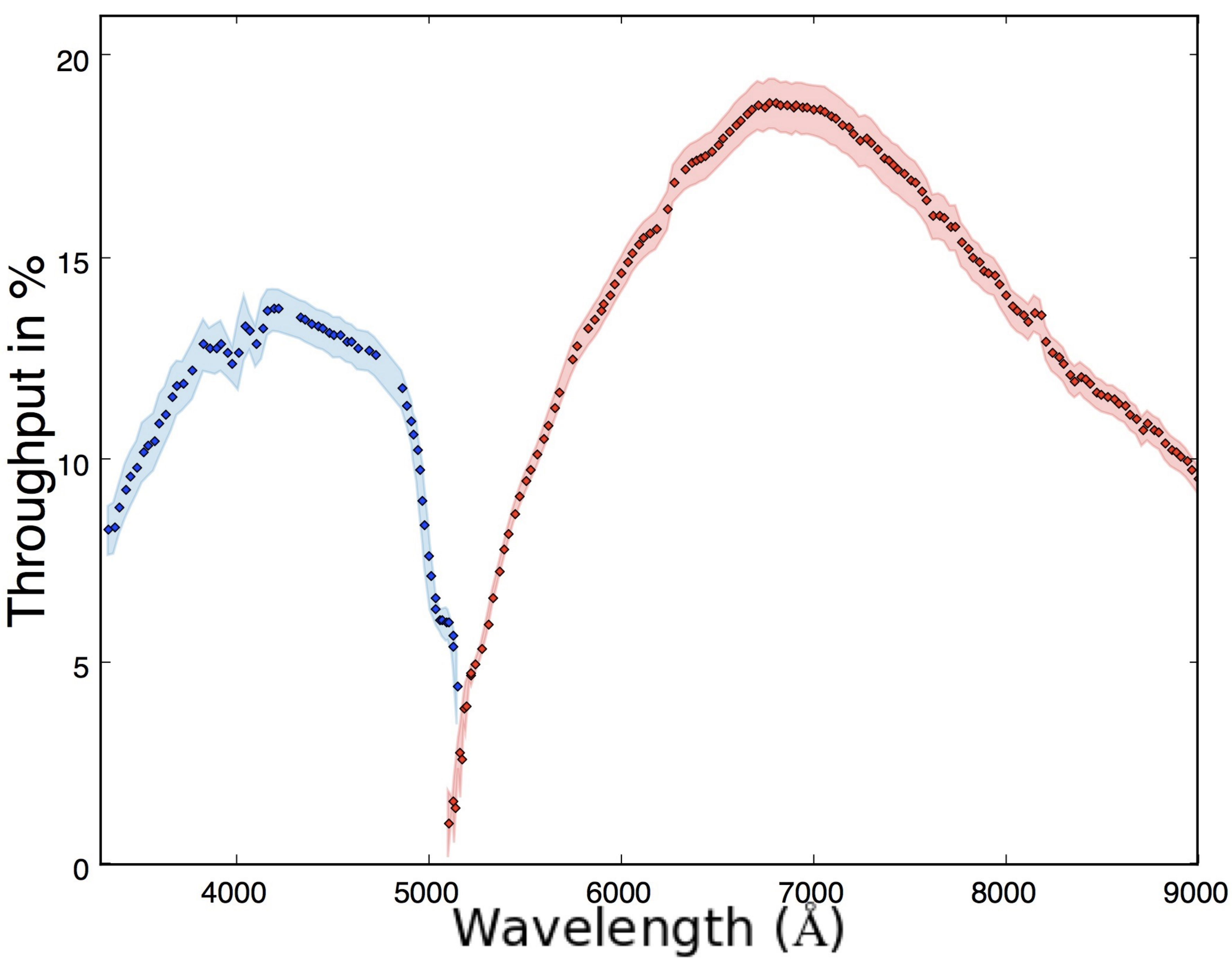}

\caption{Daytime measurements of the throughput $T_{\lambda_i,spx}$ averaged over the spaxels are shown as percentages. The blue and red circles represent the throughput measured by illuminating the blue and red spectroscopic channels respectively and the telescope with SCALA during daytime on June~4, 2015. The colored bands indicate the standard deviation between SNIFS spaxels, and are thus \emph{not} an estimate of the throughput uncertainty.}
\label{Flo:through_scala2}
\end{figure}

We have also tested performing SNIFS throughput measurements with reduced exposure times, and during daytime. 
These changes would allow daily determinations of the telescope$+$SNIFS throughput. 
On June~4, 2015 we tested this possibility by running a SCALA calibration with a wavelength sampling of 30$\,{\AA}$ and reducing the exposure times by a factor of 3.
This measurement set was performed during daytime, three days before the nighttime measurements of Figure~\ref{Flo:through_scala0}, and took about 4 hrs. The result is shown in Figure~\ref{Flo:through_scala2}.
Even though the statistical precision is lower than usual and there is some ambient light contamination, the overall result is still a smooth, high-precision measurement. Note that, as with the previous throughput curves, the shaded regions show the scatter between spaxels (not the throughput uncertainty).

Therefore, SCALA can also be used for fast estimation of throughput changes, e.g., before and after cleaning of the optics, or filter replacement.
Moreover, due to the $1^\circ$ angular extent of the SCALA calibration beam as seen at the focal plane of the telescope, such a device can also be used by other instruments mounted at the telescope having wide fields of view. 

\section{Conclusions}
\label{sec:conclusion}

We have described the SNIFS Calibration Apparatus, SCALA, built as part of a Nearby Supernova Factory effort to reach 1\% fundamental color-calibration with the SuperNova Integral Field Spectrograph. 
SCALA generates 18 collimated beams with opening angles of $1^\circ$ that produce uniform illumination across the focal plane of the telescope.
The lamp+monochromator system illuminates the telescope with monochromatic light with FWHM of 35\,{\AA}. 
The flux calibration of a NIST photodiode has been transferred to two photodiodes used as the reference system.
SCALA is meant to transfer the NIST flux calibration to the UH\,88$+$SNIFS system, and use this calibrated system to determine a NIST-traceable flux calibration of the standard star network.

SCALA development had several testing and commissioning stages.
 First, in our laboratory, we characterized the relative color responses of the individual SCALA components (six ISs, six fiber bundle arms and 18 mirrors).
Their integration into the final in-situ system results in the transmissions of each SCALA beam.
During commissioning we measured the SCALA beam transmissions again, but now using the fully assembled and mounted set-up.
 We held the reference CLAP photodiode stationary, to always monitor the same SCALA beam, and then moved the other CLAP sequentially to the 17 remaining beams.
For each one of these we observed a series of monochromatic lines, thereby obtaining the relative wavelength response of the 17 beams with respect to the reference beam.
 The normalized sum over these relative measurements provides the color of the SCALA light that illuminates SNIFS.
 Comparing this quantity with the independent component-wise measurements, we showed that we can reproduce the color trend of the total light generated by SCALA at a level better than 1\% (Section~\ref{sec:control}).
By repeating the in-situ measurements after one year, we showed how the system evolves on this timescale.
These two measurements agree very well for wavelengths above 4700\,{\AA}, but disagree at bluer wavelengths.
This disagreement is <4\% and is likely due to an evolution of the reference IS and/or fiber bundle arm with respect to the others, and is confined to wavelengths bluer than 4700\,{\AA}.
It is therefore essential to repeat the characterization of the 18 beams for these wavelengths (at a minimum) before recalibrating the standard stars.

The calibration strategy selected for SCALA allows us to perform a complete UH\,88$+$SNIFS calibration from 3200\,{\AA} to 10000\,{\AA} with steps of 30\,{\AA} in about 8\,hrs, with exposure times for each SCALA line ranging from  30\,s to 180\,s. 
We use the monochromator's shutter to control light through SCALA and we limit the maximum number of wavelengths observed per exposure to 4 for the blue channel calibration (3300 - 5000\,{\AA}) and to 10 for the red one (5000 - 10000\,{\AA}).

In the last part of the paper (Section~\ref{sec:data_take}) we have produced a throughput curve of the UH\,88$+$SNIFS system measured from nighttime observations of SCALA light.
This measurement is a combination of three different SCALA observations performed during the same night (twilight, middle of the night and sunrise).
Reobserving nine SCALA lines for each of these observations, we verified that the SCALA$+$UH\,88$+$SNIFS system is reproducible within 0.3\% above 4000\,{\AA} and 2\% below that.
We give upper limits to the main systematics (Section~\ref{sec:systematics}) affecting our measurements, which are all $<0.7$\% over the wavelength range of interest.
We show, for example, that we can efficiently remove the ambient light contamination and the variations in the dark current from the photodiodes data to better than 0.5\%, by continuously monitoring the light from SCALA before the monochromator shutter opens and after it closes.
These two measurements provide a background reconstruction that is representative of the dark current and ambient light level during the light exposure, as long as these quantities vary smoothly and on long time scales, e.g. during nighttime or in daytime when there are no clouds modulating the dome light leaks.
SCALA can, therefore, be used for daily throughput measurements during daytime. 
In order to not interfere with other daytime operations in the dome, a shorter calibration time can be achieved by either selecting a coarser sampling  or reducing the exposure time for each wavelength.

To date, one primary standard star (GD153) and several secondaries (many from the CALSPEC\footnote{\url{http://www.stsci.edu/hst/observatory/cdbs/calspec.html}} archive) have been observed with SNIFS in conjunction with SCALA observations.
The work to use the SCALA calibration described here to recalibrate these standard stars is in progress. 
Mauna Kea has very stable atmospheric conditions, in principle allowing repeatability better than a millimag \citep{mann2011}. Our typical standard star observations have $S/N > 100$ per resolution element, and the end-to-end statistical repeatability achieved with SNIFS is in the 1--2\% range. The primary limitation seems to be small differences between the true and model PSF \citep{buton_atmospheric_2013}. Most of this variation is achromatic; the color repeatability is closer to 0.5\%. Thus, observations over just a few nights will be needed in order for the statistical errors for a new standard star to become smaller than the fundamental calibration uncertainty anticipated from SCALA. 
Once the fundamental flux calibration is established, any spectrophotometric instrument -- SNIFS, STIS, \textit{Gaia} \citep{carrasco2016}, WFIRST \citep{spergel2015} -- can transfer this system to fainter stars for use by large telescopes.

\begin{acknowledgements} 

We are grateful to Kyan Schahmaneche, Laurent Le Guillou and Nicolas Regnault from LPNHE for providing the CLAP modules and support. We thank the technical staff of the University of Hawaii 2.2 m telescope. We recognize the significant cultural role of Mauna Kea within the indigenous Hawaiian community, and
we appreciate the opportunity to conduct observations from this revered site. This work was supported in part by the Director,
Office of Science, Office of High Energy Physics of the U.S. Department of Energy under Contract No. DE-AC02-
05CH11231. Support in France was provided by CNRS/IN2P3, CNRS/INSU, and PNC; LPNHE acknowledges support
from LABEX ILP, supported by French state funds managed by the ANR within the Investissements d’Avenir programme
under reference ANR-11-IDEX-0004-02. Support in Germany was provided by the German Science Foundation through TRR33 "The Dark Universe" as well as through the graduate school "Mass, Spectrum, Symmetry" GRK1504 of the Humboldt University Berlin, the Technical University Dresden and DESY Zeuthen. In China support was provided from Tsinghua University
985 grant and NSFC grant No 11173017. Some results were obtained using resources and support from the National
Energy Research Scientific Computing Center, supported by the Director, Office of Science, Office of Advanced Scientific
Computing Research of the U.S. Department of Energy under Contract No. DE-AC02- 05CH11231. We thank the Gordon \& Betty Moore Foundation for their continuing support.

\end{acknowledgements}
\bibliographystyle{aa} 
\bibliography{paper.bib}

\end{document}